\documentclass{jalc}%%  --standard mode, for preparing the paper
\usepackage{amsmath}
\usepackage{amsfonts}
\usepackage{amssymb} 
\usepackage{amscd}
\usepackage{mathrsfs}
\usepackage{mathtools}
\usepackage{indentfirst}
\usepackage{fancyhdr}
\usepackage{color}
\usepackage{tikz}
\usepackage{listings}
\usepackage{pgfplots}
\usepackage{booktabs}
\usepackage{xcolor}
\usepackage{pgf}
\usetikzlibrary{arrows,automata,positioning,fit,calc,shapes}
\usepackage{graphicx}
\usepackage{empheq}
\usepackage{caption}
\usepackage[babel]{csquotes}
\DeclareRobustCommand{\s}{s }
\newtheorem{conjecture}[theorem]{Conjecture}
\newtheorem{procedure}[theorem]{Procedure}

\begin{document}%
     %
     % Opening the front page and introducing the title:
     \title{On random primitive sets, directable NDFA\s and the generation of slowly synchronizing DFA\s}
%\title{On the randomized generation of slowly synchronizing automata: a primitive set approach}
     % The title will automatically be set with upper case letters. If 
     % you need lower case letters inside the title, then put that part 
     % into a robust command like above. 
     %
     % Footnote like things related to the paper as a whole or to
     % the title itself:
%\thanks{This paper is an example for using the `jalc' document class.}

     % Running headline if the title is too long (comment or remove
     % this command if the title is not too long, then the title will 
     % be taken as the running title):
\runningtitle{On random primitive sets, direct. NDFA\s and the generation of slowly synchr. DFA\s}
     
     % Running headline for the list of authors, the surnames are given
     % fully, the other names are abbreviated. Every name should appear
     % in small capitals (\textsc{..})
\runningauthors{\textsc{C.~Catalano}, \textsc{R. M.~Jungers}}
     
     % Alternative for running authors, if the line above is too long:
%\runningauthors{\textsc{F. Author} et~al.}

     % Now, we code all necessary data about the first author
     % first. Before doing so, we choose some suitable label
     % names, for linking this \author to his \address(es) and
     % \thanks(es):
     % `MI'   -- stands for some university department in MILANO,
     %           where the first author is affiliated,
     % `CAL'  -- stands for some CALCIO Grant Agency, supporting
     %           the first author,
     % `MIGR' -- stands for MIGRATION, since we want to announce,
     %           by the use of a \thanks (imitating an exceptional
     %           footnote in the preamble), that the first author
     %           will be wandering around the South Pole,
     % `REN'  -- stands for RENAMED, since we want to announce, by
     %           the use of a \thanksmark and a \thanks (imitating
     %           another exceptional footnote in the preamble,
     %           this time inside the argument of an \address),
     %           that the university department will be renamed.

\author[MI,CAL]{Costanza Catalano}
\address[MI]{Gran Sasso Science Institute\\
Viale Francesco Crispi 7, L'Aquila, Italy\\
  \email{costanza.catalano@gssi.it }
  %\email[Th.~Author]{tha888@dsn.uniadmi.it}
     % If the owners of email addresses could easily be identified
     % from the email addresses themselves, the optional
     % parameters could be omitted, e.g.:
  %\email{first.author@dsn.uniadmi.it}
  %\email{third.author@dsn.uniadmi.it}
}
%\thanks[UCL]{Partially supported by CALCIO Grant Agency, under the
 % project JALCPR 2016.03.30.}

     % In the same way, we code all data related to the second
     % author. Notice that we use a separate \author command for
     % each of the authors. Similarly, we use a separate \address
     % command for each of the two addresses the second author is
     % affiliated with.
     % Since no one else is linked to the \address-es and \thanks
     % presented here, a single label will do:
     % `KSC'  -- stands for data related to the author from
     %           KOSICE.
\author[CAL,MIGR]{Rapha\"{e}l M. Jungers}
\address[CAL]{ICTEAM Institute, UCLouvain\\Avenue Georges Lema\^{i}tres 4-6, Louvain-la-Neuve, Belgium\\
  \email{raphael.jungers@uclouvain.be}
     % The optional parameter in \email is not required, even
     % though the email does not display the author's name, since
     % there is only one author linked to the address in Kosice,
     % and hence there are no problems with the owner's
     % identification.
}
%\address[KSC]{Castillo de San Marco\\
%  13352 Traveling Salesman Dr.\ \#122,
%  St.\,Augustine, FL 32084,
%  USA
%}
\thanks[MIGR]{R. M. Jungers is a FNRS Research Associate. He is supported by the French Community of Belgium, the Walloon Region and the Innoviris Foundation.}

     % Now we enter data related to the last author. Since this
     % author is affiliated with the same university department as
     % the first author, we do not duplicate the same address
     % twice, putting the label `MI' will do, but we had to go
     % back and insert the email address for the last author in
     % the \address of this university department.
     % One more new label is needed, for linking the last author
     % to a separate grant support:
     % `GIO'  -- stands for some GIOCHI Grant Agency, supporting
     %           the last author.
%\author[MI,GIO]{Third Author}
%\thanks[GIO]{Supported by GIOCHI Grant Agency.}

     % Before closing the preamble by \maketitle, we put \thanks
     % commands imitating exceptional footnotes (different from
     % those thanking for a grant support):
%\thanks[MIGR]{During the period January 2009 -- June 2009, this
 % author will be wandering around the North Pole, and hence not
 % accessible at the address shown above.}
%\thanks[REN]{This department will be renamed to `Dipartimento di
 % Scienze Naturali'$\!$\@.}
     % Now we can close the preamble. Without this, no preamble
     % will be printed!
\maketitle
%
% We are through the preamble, an abstract and keywords follow.
% The keywords are separated by commas, not capitalized (not even
% the first one), without any ending period. 
\begin{abstract}
We tackle the problem of the randomized generation of slowly synchronizing deterministic automata (DFAs) by generating random primitive sets of matrices. We show that when the randomized procedure is too simple the exponent of the generated sets is $ O(n \log n) $ with high probability, thus the procedure fails to return DFAs with large reset threshold. We extend this result to random nondeterministic automata (NDFAs) by showing, in particular, that a uniformly sampled NDFA has both a
%is $ 2 $-directable and $ 3 $-directable with high probability and that the length of its shortest 
$2$-directing word and a $ 3 $-directing word of length $ O(n\log n) $ with high probability. We then present a more involved randomized algorithm that manages to generate DFAs with large reset threshold and we finally leverage this finding for exhibiting new families of DFAs with reset threshold of order $ \Omega(n^2/4) $.% found by our algorithm.
    %This document provides a sample file in which the \LaTeX2e 
    %document class \texttt{jalc} is used and explained.
    
   % document class, \LaTeX, type setting
\end{abstract}

%
%
% Put the contents of your paper here:
%
\section{Introduction}
%\subsection{Directable (non)deterministic automata}
A complete deterministic finite state automaton (DFA) is \emph{directing} or \emph{synchronizing} if it admits a word that brings the automaton from every state to the same fixed state; a word of this kind is called a \emph{directing} or \emph{synchronizing} word. More formally, a DFA is a triple $\mathcal{A}= \langle Q,\Sigma,\delta\rangle $ where $ Q $ is a finite set of states, $ \Sigma $ is a finite set of input symbols called the alphabet and $ \delta: Q\times\Sigma\rightarrow Q $ is the transition function. A synchronizing word $ w $ is a finite sequence of letters of $ Q $ for which there exists $ v\in Q$ such that $ \delta(q,w)=v $ for every $ q\in Q $, where $ \delta $ as been extended to  $ \delta: Q\times\Sigma^*\rightarrow Q $ in the usual way.
Synchronizing DFAs appear in different research fields; for example they are often used as models of error-resistant systems \cite{Epp,Chen} and in symbolic dynamics \cite{mateescu}. For a brief account on synchronizing DFAs and their other applications we refer the reader to \cite{Volk}. One of the most longstanding open problems in this field concernes the length of the shortest synchronizing word of a synchronizing DFA $ \mathcal{A} $, called the \emph{reset threshold} of the automaton and indicated by $ rt(\mathcal{A}) $:

\begin{conjecture}[(The \v{C}ern\'{y} conjecture \cite{Cerny})]\label{conj:cerny}
Any synchronizing DFA $ \mathcal{A} $ on $ n $ states has a sychronizing word of length at most $ (n-1)^2 $, so $ rt(\mathcal{A}) \leq (n-1)^2$.
\end{conjecture}
If the conjecture is true, the bound cannot be improved as there exists a family of automata having reset threshold of exactly $ (n-1)^2 $, known as the \v{C}ern\'{y}'s automata \cite{Cerny}. Despite great effort, for long time the best upper bound known for the reset threshold of an $ n $-state synchronizing DFA was $ (n^3-n)/6 $ \cite{Frankl,Pin}, recently improved to  $(15617 n^3 + 7500 n^2 + 9375 n - 31250)/93750$ \cite{Szykula}.
 %$ (n^3-n)/6 $, obtained more than 30 years ago by J.-E. Pin and P. Frankl \cite{Frankl,Pin} and recently improved to 
 % $(15617 n^3 + 7500 n^2 + 9375 n - 31250)/93750$, recently obtained by Szyku{\l}a in \cite{Szykula} and thereby beating the 30 years-standing upper bound of $ (n^3-n)/6 $ found by Pin and Frankl \cite{Frankl,Pin}. 
Exhaustive search has confirmed the conjecture for small values of $ n $ \cite{SlowAutom, BondtDon} while quadratic upper bounds have been obtained for certain classes of DFAs \cite{Beal, Babai, Kari, Rystsov, Volkov2007}. The search for synchronizing DFAs attaining quadratic reset threshold (called \textit{extremal} or \emph{slowly synchronizing} automata) has been the subject of several contributions in recent years, partially due to the fact that they are hard to detect %, partially due to the fact that search space is wide and calculating the reset threshold is computationally hard %despite deciding whether an automaton is synchronizing is polynomial in time 
%(see \cite{Epp,Olsch}); 
and few families are known (see \cite{SlowAutom,BondtDon,GusevSzikulaDzyga,Szykula2015,Szykula2016,Babai} for examples). %These results are partly summarized in Table \ref{table}.
%Slowly synchronizing automata are hard to detect %, partially due to the fact that search space is wide and calculating the reset threshold is computationally hard %despite deciding whether an automaton is synchronizing is polynomial in time 
%(see \cite{Epp,Olsch}); 
%and few families are known (see 
The great majority of these extremal DFAs is two-letter and has a quite regular structure; in particular, the action of their letters is very much similar to the ones of \v{C}ern\'{y}'s. % $ a $ and $ b $. %$ \mathcal{C}(n)=\lbrace a,b\rbrace $ 
%, where $ a $ is the cycle over $ n $ vertices and $ b $ the letter that fixes all the vertices but one, which is mapped to the same vertex as done by $ a $; indeed all these families present a letter that is a cycle over $ n $ vertices and the other letters have an action similar to the one of letter $ b $. 
It is natural to wonder whether a randomized procedure to generate automata could obtain less structured synchronizing DFAs with possibly larger reset thresholds; this approach can be rooted back to the 60s with Erd\H{o}s and his \emph{Probabilistic Method}, where the existence of a structure with certain desired properties is proved by defining a suitable probabilistic space in which to embed the problem 
(for an account on the probabilistic method we refer the reader to \cite{AlonSpencer}). This randomized procedure cannot be too simple: indeed, Berlinkov \cite{Berl} and Nicaud \cite{Nicaud} showed that an uniformly generated 2-letter DFA is synchronizing with high probability (i.e.\ the probability that it is synchronizing tends to $ 1 $ as the number of states $ n $ tends to infinity) and it also has a synchronizing word of length $ O(n\log^3 n) $ with high probability. It follows that:
%two uniformly and independently sampled binary row-stochastic matrices form a synchronizing automaton with reset threshold of order $ O(n\log^3 n) $ with high probability (i.e.\ the probability that they form a synchronizing automaton with reset threshold of $ O(n\log^3 n) $ tends to $ 1 $ as the matrix dimension tends to infinity). %Therefore, a slowly syncrhonizing automaton is almost surely never generated by an uniform random generation.   
\begin{itemize}
\item slowly synchronizing DFAs are almost surely never generated by a uniform distribution;
%Slowly synchronizing automata cannot be generated by a mere uniform randomized procedure;
\item synchronizing DFAs with more than two letters that need every letter to synchronize (called \emph{proper} automata) are hard to find, as usually two letters are enough to make the automaton be synchronizing. As \emph{proper} automata do not appear often in the literature, they are especially of interest since the behavior of their reset threshold is still unclear.
\end{itemize}
With this in mind, we decided to approach the randomized generation of (slowly synchronizing) DFAs by enforcing them to be \emph{proper}; to accomplished this, we make use of the concept of \emph{primitive sets}, described in the next paragraph.\\
The notion of synchronization can be generalized to nondeterministic finite automata (NDFA) in several ways (see for example \cite{Imreh}); here we will focus on the $ 2 $\emph{-directability} and the $ 3 $\emph{-directability} properties, that we now describe. 
A NDFA is defined as a triple $ \mathcal{N}=\langle Q,\Sigma,\delta\rangle $ where $ Q $ is a finite set of states, $ \Sigma $ is a finite set of input symbols and $ \delta\subseteq Q\times\Sigma\times Q $ is the transition function; this time the transition from one state to another by a letter may be not defined or not uniquely defined. An NDFA is $  2$\emph{-directable} if there exists a word $ w $ (called a \textit{$ 2 $-directing} word) such that $\delta(q,w)=\delta(p,w) $ for every $ p,q\in Q $, where $ \delta $ has been extended to $ \delta\subseteq Q\times\Sigma^*\times Q $ in the usual way; in other words, an NDFA is $ 2 $-directable if the set of states that can be reached by applying the word $ w $ is independent of the initial state. An NDFA is $  3$\emph{-directable} if there exist $ v\in Q $ and a word $ w $ (called a \emph{$ 3 $-directing word}) such that $ v\in \delta(q,w) $ for every $ q\in Q $; in other words, an NDFA is $ 3 $-directable if there exists a state that is reachable from any other state by applying the word $ w $. Similarly to synchronizing DFAs, we can define $  d_2(\mathcal{N}) $ the length of the shortest $ 2 $-directable word of an NDFA $ \mathcal{N} $, $  d_3(\mathcal{N}) $ the length of its shortest $ 3 $-directable word and $ d_i(n)=\max\lbrace d_i(\mathcal{N}): \mathcal{N} \text{ is an $i$-directable NFDA on $ n $ states} \rbrace $, for $ i=2,3 $. It is known that $ d_2(n)=\Theta(2^n) $ \cite{GAZDAG,Burkhard}, $ d_3(n)=O( 4^{n/3}n^2) $ \cite{GAZDAG} and $ d_3(n)=\Omega(3^{n/3}) $ \cite{Martyugin} but, to the best of our knowledge, it is still not clear what is the \emph{average} behavior of an NDFA; more precisely, we wonder what is the probability that a random NDFA $ \mathcal{N} $ is $ 2 $- or $ 3 $-directable as $ n\rightarrow\infty $ and what is the \emph{expected} magnitude of $  d_2(\mathcal{N}) $ and $  d_3(\mathcal{N}) $. Our primitive sets approach will also provide an answer this question.

\subsection{The primitive set approach}
A finite set of nonnegative matrices $ \mathcal{M} = \lbrace M_1, \dots ,M_m\rbrace $ is called \textit{primitive} if there exists a product $M=M_{i_1}\cdots M_{i_l}>0 $ entrywise, for some $ i_1,\dots ,i_l\in \lbrace 1,\dots ,m\rbrace $; %A finite set of nonnegative %\footnote{A matrix is \textit{nonnegative} if it has nonnegative entries.} 
%matrices is called \emph{primitive} if there exists a product of these matrices with all positive entries; 
in this case $ M $ is called a \emph{positive} product. The \textit{exponent} of a primitive set ($ exp(\mathcal{M})$) is the length of its shortest positive product.
The concept of primitive set has been introduced by Protasov and Voynov in \cite{ProtVoyn} as an extension of the notion of primitive matrix due to Frobenius in 1912 and has found application in different fields as in stochastic switching systems \cite{ProtJung}, consensus for discrete-time multi-agent systems \cite{Pierre}, time-inhomogeneous Markov chain \cite{Hart} and, finally, automata theory \cite{BlonJung,GerenGusJung}.
\begin{remark}
Any NDFA $\mathcal{N}=\langle Q,\Sigma,\delta\rangle$ with $ Q=\lbrace 1,\dots,n\rbrace $ and $ \Sigma=\lbrace a_1,\dots ,a_m\rbrace $ can be uniquely represented by the matrix set $ \lbrace A_1,\dots ,A_m\rbrace $ where, for all $ i=1,\dots ,m $, $ A_i[l,k]=1 $ if $ k\in\delta(l,a_i) $, $ A_i[l,k]=0 $ otherwise. Equivalently, any set of binary\footnote{A matrix is \textit{binary} if it has entries in $ \lbrace 0,1\rbrace $.} matrices is an NDFA. In this context, $\mathcal{N} $ is $ 2 $-directable iff there exists a product $ A=A_{j_1}\cdots A_{j_s} $ for some $ j_1,\dots ,j_s\in \lbrace 1,\dots ,m\rbrace $ such that every column of $ A $ is either entrywise positive, or entrywise equal to $ 0 $. %and a $ J\subseteq Q $ such that, for every column $ c $ of $ A $, either the support\footnote{The \textit{support} of a nonnegative vector is the set of indices of its positive entries.} of $ c $ is equal to $ J $ or it is the empty set. 
Similarly, $\mathcal{N} $ is $ 3 $-directable iff there exists a product $ A_{j_1}\cdots A_{j_s} $ with an entrywise positive column, for some $ j_1,\dots ,j_s\in \lbrace 1,\dots ,m\rbrace $. Note that in case of DFAs, the matrices $ A_1,\dots ,A_m $ are row-stochastic\footnote{A nonnegative matrix is \textit{row-stochastic} if the entries of each row sum up to $ 1 $. In the case of a letter of a DFA, it means that each row of the matrix has exactly one $ 1 $.} and the DFA is synchronizing iff it admits a product with an all-ones column\footnote{A column whose entries are all equal to $ 1 $.} (while all the other columns are made of zeros).
\end{remark}
In the rest of the paper we will mostly use this matrix representation of DFAs and NDFAs.
It is clear that a primitive set $ \mathcal{N} $ of binary matrices is both a $ 2 $-directable and a $ 3 $-directable NDFA and it holds that 
\begin{equation}\label{eq:d1d2}
\max\lbrace d_2(\mathcal{N}), d_3(\mathcal{N})\rbrace\leq exp(\mathcal{N}).
\end{equation}
A less obvious connection between synchronizing DFAs and primitive sets is due to the following Definition \ref{def:assoc_autom} and Theorem \ref{thm:autom_matrix}; we call a nonnegative matrix \emph{NZ} if it has at least one positive entry in every row and in every column.
%A novel approach to the study of synchronizing automata was being developed by Blondel et.\ al.\ in \cite{BlonJung} and then further exploited by Gerencs\'{e}r et.\ al.\ in \cite{GerenGusJung}: this approach is based on the concept of \textit{primitive set of matrices}, i.e. a finite set of nonnegative\footnote{A matrix is \textit{nonnegative} if it has nonnegative entries.} matrices that admits a product, with repetitions allowed, with all positive entries; a product of this kind is called a \emph{positive} product. This notion was introduced for the first time by Protasov and Voynov in \cite{ProtVoyn} as an extension of the notion of primitive matrix due to Frobenius in 1912 and has found application in different fields as in stochastic switching systems \cite{ProtJung}, consensus for discrete-time multi-agent systems \cite{Pierre} and time-inhomogeneous Markov chain \cite{Hart}. 
%One of the connections between primitive sets and automata is established by Theorem \ref{thm:autom_matrix} presented below, which summarizes two results proved by Blondel et.\ al.\ (\cite{BlonJung}, Theorems 16-17) and a result proved by Gerencs\'{e}r et.\ al.\ (\cite{GerenGusJung}, Theorem 8); in the following we call the \textit{exponent} of a primitive set $ \mathcal{M} $ (denoted by $ exp(\mathcal{M})$) the length of its shortest positive product and a nonnegative matrix is said to be \emph{NZ} if it has neither zero-rows nor zero-columns\footnote{Thus a NZ-matrix must have a positive entry in every row and in every column.}.

\begin{definition}\label{def:assoc_autom}
 Let $ \mathcal{M}$ be a set of binary NZ-matrices. % and, for any $ k=1,\dots ,m $, let $ \tilde{M}_k $ be the binary matrix such that $ \tilde{M}_k(i,j)=1 $ if and only if $ M_k(i,j)>0 $, for every $ i,j $. 
The \emph{DFA associated to} the set $ \mathcal{M} $ is the automaton \[ \mathcal{A}(\mathcal{M})=\lbrace A: A \text{ is  binary row-stochastic, }\exists M\in\mathcal{M} \,\,s.t.\,\, A[i,j]\leq M[i,j], \forall i,j  \rbrace. \]
 %whose letters are all the binary row-stochastic matrices that are entrywise not greater than at least one matrix in $\mathcal{M}$.
\end{definition}
For an example of a set $ \mathcal{M}$ and its associated DFA $ \mathcal{A}(\mathcal{M}) $ see Example \ref{ex}.
\begin{theorem}[(\cite{BlonJung}, Theorems 16-17 and \cite{GerenGusJung}, Theorem 8)]\label{thm:autom_matrix}$  $\\
Let $ \mathcal{M}\!=\!\lbrace M_1,\dots ,M_m\rbrace $ be a set of $ n\times n $ binary NZ-matrices and let $ \mathcal{M}^T=\lbrace M^T_1,\dots ,M^T_m\rbrace  $. The set $ \mathcal{M} $ is primitive if and only if $ \mathcal{A}(\mathcal{M})$ (equiv. $ \mathcal{A}(\mathcal{M}^T)$) is synchronizing. If $\mathcal{M}  $ is primitive, it also holds that:
 %there exists two $ n $-states synchronizing automata $ \mathcal{A} $ and $ \mathcal{A}' $ such that:
\begin{equation}\label{eq:thmauotm_mat}
\!\!\!\!\max \Bigl\lbrace rt\bigl(\mathcal{A}(\mathcal{M})\bigr),rt\bigl(\mathcal{A}(\mathcal{M}^T)\bigr)\Bigr\rbrace\leq exp(\mathcal{M}) \leq rt\bigl(\mathcal{A}(\mathcal{M})\bigr)+rt\bigl(\mathcal{A}(\mathcal{M}^T)\bigr)+n-1.
\end{equation}
%where $ \mathcal{A} $ is the set of all the binary row-stochastic matrices that are entrywise smaller than at least one matrix in $ \mathcal{M} $ and $ \mathcal{A}' $ is the set of all the binary row-stochastic matrices that are entrywise smaller than at least one matrix in $ \mathcal{M}^T=\lbrace M^T_1,\dots ,M^T_m\rbrace  $. 
%Furthermore $ \mathcal{M} $ is primitive if and only if $ \mathcal{A} $ (equiv. $ \mathcal{A}' $) is synchronizing.
\end{theorem}

\begin{example}\label{ex}
We here present a primitive set $ \mathcal{M}$ and the DFAs $ \mathcal{A}(\mathcal{M}) $ and $ \mathcal{A}(\mathcal{M}^T) $.% in both their matrix and graph representation (Figure \ref{fig:twoautom}).
%associated, respectively, to $ \mathcal{M} $ and $ \mathcal{M}^T $ 
\begin{align*}
\mathcal{M}\!&=\!\left\lbrace  
\left( \begin{smallmatrix} 0 & 1&0 \\ 1&0&0 \\  0&0&1 \end{smallmatrix}\right) , \left( \begin{smallmatrix} 1 & 0&1 \\  0&0&1 \\ 0&1 & 0 \end{smallmatrix}\right) \right\rbrace ,  &\mathcal{A}(\mathcal{M})\!&=\!\left\lbrace  
\left( \begin{smallmatrix} 0 & 1&0 \\ 1&0&0 \\  0&0&1 \end{smallmatrix}\right) , \left( \begin{smallmatrix} 1 & 0&0 \\  0&0&1 \\ 0&1 & 0 \end{smallmatrix}\right) ,\left(  \begin{smallmatrix} 0 & 0&1 \\  0&0&1 \\ 0&1 & 0 \end{smallmatrix}\right) \right\rbrace \!=\! \lbrace a,b,c\rbrace ,\\
\mathcal{M}^T\!&=\!\left\lbrace  
\left( \begin{smallmatrix} 0 & 1&0 \\ 1&0&0 \\  0&0&1 \end{smallmatrix}\right) , \left( \begin{smallmatrix} 1 & 0&0 \\  0&0&1 \\ 1&1 & 0 \end{smallmatrix}\right) \right\rbrace, &\mathcal{A}(\mathcal{M}^T)\!&=\!\left\lbrace  
\left( \begin{smallmatrix} 0 & 1&0 \\ 1&0&0 \\  0&0&1 \end{smallmatrix}\right) ,\left(  \begin{smallmatrix} 1 & 0&0 \\  0&0&1 \\ 0&1 & 0 \end{smallmatrix}\right) ,\left(  \begin{smallmatrix} 1 & 0&0 \\  0&0&1 \\ 1&0 & 0 \end{smallmatrix}\right) \right\rbrace \!=\! \lbrace a,b,c'\rbrace.
\end{align*}
One can verify that $ exp(\mathcal{M})=8=exp(\mathcal{M}^T) $, $ rt\bigl(\mathcal{A}(\mathcal{M})\bigr)=4 $ and $ rt\bigl(\mathcal{A}(\mathcal{M}^T)\bigr)=2 $.
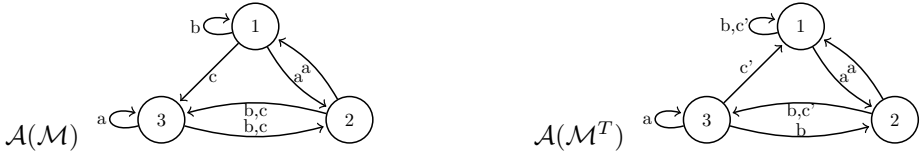
\begin{figure}[h!]
$ \mathcal{A}(\mathcal{M})\,\, $
\begin{tikzpicture}[shorten >=1pt,node distance=2.5cm,on grid,auto,scale=0.7,transform shape,inner sep=0pt,bend angle=15,line width=0.2mm]

%node[shape = circle split, draw, line width = 1pt,
         % minimum size = 10mm, inner sep = 0mm, font = \sffamily\large,
          %rotate=30]

\node[state]    (q_1) {1 };
 			\node[state]    (q_2) [ below right=of q_1] {2};
 			\node[state]          (q_3) [below left=of q_1] {3};

 			\path[->] (q_1) edge	[loop left]	node  {b$ \,$} (q_2)
 			(q_1) edge	[bend right]	 node  {a} (q_2)
 			(q_2) edge [bend right]	node  {a} (q_1)
 			(q_3) edge [loop left]		node  {a$ \,$ } ()
 			(q_2) edge [bend right]		node  {b,c} (q_3)
 			(q_3) edge [bend right]		node  {b,c} (q_2)
 			(q_1) edge 		node  {c} (q_3)
 			;  
\end{tikzpicture}$ \qquad\qquad\qquad\mathcal{A}(\mathcal{M}^T)\,\, $
\begin{tikzpicture}[shorten >=1pt,node distance=2.5cm,on grid,auto,scale=0.7,transform shape,inner sep=0pt,bend angle=15,line width=0.2mm]

%node[shape = circle split, draw, line width = 1pt,
         % minimum size = 10mm, inner sep = 0mm, font = \sffamily\large,
          %rotate=30]

\node[state]    (q_1) {1 };
 			\node[state]    (q_2) [ below right=of q_1] {2};
 			\node[state]          (q_3) [below left=of q_1] {3};

 			\path[->] (q_1) edge	[loop left]	node  {b,c'} (q_2)
 			(q_1) edge	[bend right]	 node  {a} (q_2)
 			(q_2) edge [bend right]	node  {a} (q_1)
 			(q_3) edge [loop left]		node  {a$ \,$} ()
 			(q_2) edge [bend right]		node  {b,c'} (q_3)
 			(q_3) edge [bend right]		node  {b} (q_2)
 			(q_3) edge 		node  {c'} (q_1)
 			;  
\end{tikzpicture}
\caption{The automata $ \mathcal{A}(\mathcal{M}) $ and $ \mathcal{A}(\mathcal{M}^T) $ of Example \ref{ex}.}\label{fig:twoautom}
\end{figure}
\end{example}
Theorem \ref{thm:autom_matrix} more generally holds for any set of NZ-matrices with nonnegative entries, due to the fact that the property of being primitive is not influenced by the actual values of the positive entries of the matrices of the set. In this case the automaton $ \mathcal{A}(\mathcal{M}) $ of Definition \ref{def:assoc_autom} should be defined as $ \mathcal{A}(\mathcal{M})=\lbrace A: A \text{ is  binary row-stochastic, }\exists\,\, M\!\in\!\mathcal{M} \,\,s.t.\,\, \forall\, i,j,\, M[i,j]\!=\!0 \,\Rightarrow \, A[i,j]\!=\!0   \rbrace$.
%we should add to Definition \ref{def:assoc_autom} the request of setting to $ 1 $ all the positive entries of the matrices of $ \mathcal{M} $ before building $ \mathcal{A}(\mathcal{M}) $. 
Equation (\ref{eq:thmauotm_mat}) shows that primitive sets can be used for generating synchronizing DFAs; a primitive set with large exponent implies that the associated DFA has large reset threshold. In particular, an NZ-primitive set with exponent greater than $ 2(n-1)^2-n+1 $ would disprove the \v{C}ern\'{y} conjecture. 
%Primitivity of NZ-sets is easily checked by the Protasov-Voynov algorithm (\cite{ProtVoyn}, Proposition 2), while the problem in the general case is NP-hard \cite{GerenGusJung}.
%On the other hand, the upper bounds on the automata reset threshold mentioned before imply that $ exp(\mathcal{M})\!=\!O(n^3) $.
%QYUQYUQYU

%\begin{enumerate}
%\item [\textbf{Q1}] Is it possible to randomly generate NZ primitive sets with large exponent %thus leading to automata with large reset thresholds?
%\end{enumerate}
\subsection{Our contribution}
The present article aims to answer the two following questions:
\begin{enumerate} 
\item [\textbf{Q1}] Is it possible to randomly generate NZ-primitive sets with large exponent thus leading to synchronizing DFAs with large reset threshold?
\item [\textbf{Q2}] What is the probability that a random NDFA is $ 2 $- or $ 3 $-directable? What is the expected length of its shortest $ 2 $-directing and $ 3 $-directing words?  
\end{enumerate}
In Section \ref{sec:whp} we give a negative answer to question \textbf{Q1} in case the randomized generation is too simple and we answer question \textbf{Q2}. In Subsection \ref{subsec:randperturb} we show that a uniformly generated perturbed permutation set (see Definition \ref{defn:pertpermset}, Section 3) is primitive and has exponent of order $ O(n\log n) $ with high probability, which implies that its associated DFA has reset threshold of order $ O(n\log n) $ with high probability. This result leads us to the main theorem of Section \ref{sec:whp}, presented in Subsection \ref{subsec:randbin}: we show that a random binary set of $ n\times n $ matrices generated by setting each entry of each matrix to $ 1 $ with probability $ p $ and to $ 0 $ with probability $ 1-p $, independently of each others, is primitive and has exponent of order $ O(n\log n) $
 with high probability when $ p\geq (1+\alpha)(\log n+c)/n $ for some $ c\in\mathbb{R} $ and $ \alpha>0 $, while it is almost surely never primitive when $ p\leq (1-\alpha)(\log n+c)/n $ for some $ c\in\mathbb{R} $ and $ \alpha>0 $. %This implies that a randpm binary set generated by the uniform distribution ($ p=1/2 $) is pirmitive with high probability. 
In the case $ p=(\log n+c)/n $ for some $ c\in\mathbb{R} $, the set exhibits an intermediate behavior and we show that its exponent is of order $ O(n\log^3n) $ with high probability under some conditions. In other words, $p=(\log n+c)/n $ is a \emph{sharp threshold} for the property of these sets to be primitive and this result show that their associated DFAs have small reset threshold most of the times.
As corollaries, in Subsection \ref{subsec:randNDFA} we show that any NDFA randomly generated as described above is $ 2 $-directable and has a $ 2 $-directing word of length $O(n\log n)  $ with high probability when $ p\geq (1+\alpha)(\log n+c)/n $ for some $ c\in\mathbb{R} $ and $ \alpha>0 $, and that the $ 3 $-directability property of these sets has the same \emph{threshold} behavior described for primitivity. In particular, a random NDFA generated according to the uniform distribution ($ p=1/2 $) has both a $ 2 $-directing word and a $ 3 $-directing word of length $ O(n\log n) $ with high probability.
 %for the $ 3 $-directable property, i.e.\ if $ p(n)\geq (1+\alpha)(\log n+c)/n $ (asymptotically), the random NDFA is $ 3 $-directable and $ d_3(\mathcal{N})=O(n\log n) $ with high probability, if $ p(n)\leq (1-\alpha)(\log n+c)/n $ (asymptotically) the set is \emph{not} $ 2 $-directable with high probability and when $ p(n)=(\log n+c)/n $ (asymptotically) the probability that the NDFA is $ 3 $-directable is strictly between $ 0 $ and $ 1 $ as $ n\rightarrow\infty $ and $d_3(\mathcal{N})= O(n\log^3n) $ with high probability under the condition that all the matrice are NZ.
 \\
In Section \ref{sec:algorithm} we present a more involved randomized algorithm that manages to generate NZ-primitive sets with quadratic exponent, thus providing a positive answer to question \textbf{Q1}. The algorithm generates proper\footnote{A primitive set is \textit{proper} if it needs all its matrices to be primitive.} primitive perturbed permutation sets (see Definition \ref{defn:pertpermset}) of cardinality greater than two by exploiting a combinatorial characterization theorem of NZ-primitive sets (Theorem \ref{thmProt}, Section \ref{sec:def}), and from them we obtain proper synchronizing DFAs with more than two letters. To the best of our knowledge, this is the first time where a constructive procedure for finding proper synchronizing DFAs is presented. Finally, in Section \ref{sec:famaut} we present the new families of slowly synchronizing automata found by our algorithm: they are $ 3 $-letter proper synchronizing DFAs that do not resemble the \v{C}ern\'{y}'s family and with reset threshold of order $ \Omega(n^2/4) $. This last result improves the state of the art in the direction initiated by Gonze et.\ al.\ in \cite{Gonze}: they prove that the diameter of the square graph (see Definition \ref{def:sg}, Section \ref{sec:numexp}) of any $ n $-state DFA made of $ m\!\geq\! 2 $ permutation matrices is lower bounded by $ n^2/4+o(n^2) $ when $ n $ is odd. We prove that this lower bound holds for any $ n $ and any $ n $-state \emph{synchronizing} DFA containing $ m\!\geq \! 2 $ permutation matrices. %\textcolor{blue}{Our families also generalize the slowly eulerian synchronizing automata presented in \cite{Szykula2016}}.

\section{Definitions and notation}\label{sec:def}
In this section we briefly go through some definitions and results that will be needed in the rest of the paper.\\
Sometimes we will refer to synchronizing deterministic finite automata just as \emph{synchronizing automata}; when we will consider \emph{non}deterministic finite automata it will always be specified.
We indicate with $ [n] $ the set $ \lbrace 1,\dots ,n\rbrace $ and with $ S_n $ the set of permutations over $ n$ elements; with a slight abuse of notation $ S_n $ will also denote the set of the $ n\times n $ permutation matrices, where a \emph{permutation matrix} is a binary matrix having exactly one $ 1 $ in every row and in every column. We indicate with $ \mathbb{I}_{i,j} $ the matrix 
such that $  \mathbb{I}_{i,j}[i,j]=1  $ and all the other entries are equal to $ 0 $; $ M^T $ denotes the transpose of a matrix $ M $. The set of all the binary row-stochastic matrices of size $ n\times n $ is indicated by $ \mathcal{R}_n $, $ \mathcal{C}_n:=\lbrace R^T: R\in \mathcal{R}_n \rbrace$ is the set of all the binary column-stochastic matrices and $ \mathcal{NZ} $ represents the set of all the binary NZ-matrices. \\
Given two sequences $ \lbrace a_n\rbrace,\lbrace b_n\rbrace $, $ n\in\mathbb{N} $, we say that $ a_n=O(b_n) $ if there exist $ C>0 $ and $ N\in\mathbb{N} $ such that $ a_n\leq Cb_n $ for every $ n>N $, that $ a_n=\Omega(b_n) $ if there exist $ C>0 $ and $ N\in\mathbb{N} $ such that $ a_n\geq Cb_n $ for every $ n>N $ and that $ a_n=\Theta(b_n) $ if $ a_n=O(b_n) $ and  $ a_n=\Omega(b_n) $. If $ \mathbb{P} $ is a probability distribution over a finite space $ \Omega $ and $ A,B\subset\Omega $ two events, we indicate with $ \mathbb{P}(A|B) $ the conditional probability of $ A $ given $ B $.\\
We remind that a matrix $ M $ is \emph{irreducible} if there does not exist a permutation matrix $ P $ such  that $ PMP^T $ is block-triangular; a set $ \lbrace M_1,\dots ,M_m\rbrace $ is said to be \emph{irreducible} if the matrix $ \sum_{i=1}^m M_i $ is irreducible. The \textit{directed graph associated to} an $ n\!\times \! n $ nonnegative matrix $ M $ is the digraph $ D_M $ on $ n $ vertices with a edge from $ i $ to $ j $ iff $ M[i,j]>0 $. A matrix $ M $ is irreducible if and only if $ D_M $ is strongly connected, i.e.\ if and only if there exists a directed path between any two given vertices in $ D_M $.
%A set of $ m $ nonnegative matrices $ \lbrace M_1, \dots ,M_m\rbrace $ is \textit{primitive} if there exists a product $ M_{i_1}\cdots M_{i_l}>0 $ entrywise, for $ i_1,\dots ,i_l\in [m] $. The length of the shortest of these products is called the \textit{exponent} of the set. 
Irreducibility is a necessary but not sufficient condition for a matrix set to be primitive (\cite{ProtVoyn}, Section 1). 
%A matrix is said to be NZ if it has no zero-rows nor zero-columns. %Since the concept of primitivity involves just the positions of the positive entries in the matrices and not their actual values, without loss of generality we restrict ourselves to binary matrices, where the product between two binary matrices $ A $ and $ B $ is defined by setting $ AB(i,j)=1 $ if $ \sum_{k}A(i,k)B(k,j)>0 $ and $ AB(i,j)=0 $ otherwise. This will be assumed for the rest of the paper. 
Primitive sets of NZ-matrices can be characterized as follows:
\begin{definition}\label{def2}
Let $ \Omega=\dot{\bigcup}^k_{l=1} \Omega_l $ be a partition of $ [n] $ with $ k\geq 2 $. We say that an $ n\times n $ matrix $ M $ has a \textit{block-permutation structure on the partition} $ \Omega $ if there exists a permutation $ \sigma\!\in\!S_k $ such that $\forall \,l\!=\!1,\dots ,k  $ and $\forall\, i\!\in\! \Omega_l $, if $M[i,j]>0  $ then $j\in \Omega_{\sigma(l)}$. We say that a set of matrices \textit{has a block-permutation structure} if there exists a partition on which \textit{all} the matrices of the set have a block-permutation structure.
\end{definition}
\begin{theorem}[(\cite{ProtVoyn}, Theorem 1)]\label{thmProt}
An irreducible set of NZ-matrices is \emph{not} primitive if and only if the set has a block-permutation structure.
%there exists a partition $\Omega$ of $ [n] $ on which \emph{every} matrix of the set has a block-permutation structure.
\end{theorem}
  %; if this is not the case, we say that the set has no block-permutation structure.\\
 %(see for example \cite{Hart}). 
%A reducible set cannot be primitive. %as, after a suitable permutation of the canonical basis, all the matrices share the same block upper-triangular form. 
%We end this section with the last definition and our first observation (Proposition \ref{cor:gonze}).
%first little result concerning the types of block-permutation structures that some irreducible sets can admit:
We say that a matrix $ A $ \textit{dominates} a matrix $ B $ ($ A\geq B $) if $ A[i,j]\geq B[i,j],\,\,\forall \, i,j $.

\begin{proposition}\label{cor:gonze}
Consider an irreducible set $ \lbrace M_1,\dots ,M_m\rbrace $ in which every matrix dominates a permutation matrix. If the set has a block-permutation structure, then all the blocks of the partition must have the same size.
\end{proposition}

\begin{proof} 
Let $ Q_i$ be the permutation matrix dominated by $ M_i $; if $ M_i $ has a block-permutation structure on a given partition, so does $ Q_i $ on the same partition. Theorem 2 in \cite{Gonze} states that if a set of permutation matrices has a block-permutation structure then all the blocks of the partition must have the same size, so we conclude. 
\end{proof}

\section{Primitivity and small exponent with high probability}\label{sec:whp}
\subsection{Random perturbed permutation sets}\label{subsec:randperturb}
In this section we focus on \emph{perturbed permutation} sets (see the following definition) and we show that in case of uniform distribution these sets have small exponent most of the times, which implies that their associated DFAs (see Definition \ref{def:assoc_autom}) have almost surely small reset threshold.

\begin{definition}\label{defn:pertpermset}
A  \emph{perturbed permutation} set is a matrix set made of permutation matrices where a $ 0 $-entry of one of the matrices is changed into a $ 1 $.
\end{definition}
Perturbed permutation sets are particularly of interest as they have the following properties:
\begin{itemize}
\item [-]they have the least number of positive entries that an NZ-primitive set can have, which intuitively should lead to sets with large exponent;
\item [-]their associated DFAs are easily computable;% and%. It is made of permutation letters and a letter of rank $ n\!-\! 1 $ and 
%its alphabet size is just one unit more than the cardinality of $ \mathcal{M} $;
\item [-]if they are primitive and proper, their associated DFAs are synchronizing and proper (or they can be made proper by removing one known letter, as shown in Section \ref{sec:algorithm}, Proposition \ref{prop:minsetsautom}).
%or automata $ \bar{\mathcal{A}} $ and $ \bar{\mathcal{A}}' $ obtained by removing one (known) letter from $ \mathcal{A} $ and $ \mathcal{A}' $ respectively; 
%\item [-]primitivity is easily checked by the Protasov-Voynov algorithm (\cite{ProtVoyn}, Proposition 2), and primitivity of $ \mathcal{M} $ assures that $ \mathcal{A}(\mathcal{M}) $ is synchronizing (Theorem \ref{thm:autom_matrix}).
\end{itemize}
For these reasons they will play a significant role in the randomized generation of slowly synchronizing automata in Section \ref{sec:algorithm}.\\
We call \textit{random perturbed permutation set} a perturbed permutation set of $ m \!\geq\! 2$ matrices
constructed with the following randomized procedure:
\begin{procedure}\label{proc}
\begin{enumerate}
\item $ m $ permutation matrices $ \lbrace P_1,\dots ,P_m\rbrace $ are sampled independently and uniformly from the set $ S_n $;
\item a matrix $ P_i $ is uniformly chosen from the set $ \lbrace P_1,\dots ,P_m\rbrace $ and one of its $ 0 $-entry is uniformly selected among its $ 0 $-entries and changed into a $ 1 $. It becomes then a perturbed permutation matrix $ \bar{P}_i $;
\item The final set is the set $ \lbrace P_1,\dots ,P_{i-1},\bar{P}_i, P_{i+1},\dots ,P_m\rbrace $.
\end{enumerate}
\end{procedure}
This procedure is equivalent to choosing independently and uniformly $ m-1 $ permutation matrices from $ S_n $ and one \textit{perturbed} permutation matrix from $ \bar{S}_n=\lbrace\bar{P}: \bar{P}\!=\!P+\mathbb{I}_{i,j}, \,P\!\in\! S_n,\,\exists\, i'\neq i : P[i',j]=1 \rbrace $, the set of the perturbed permutation matrices. 
We say that a property $ X $ holds for a random matrix set with \textit{high probability} if the probability that property $ X $ holds tends to $ 1 $ as the matrix dimension $ n $ tends to infinity. 
%We are now ready to state and prove the main result of this section; 
\begin{theorem}\label{Thm:primperm}
With high probability a perturbed permutation set constructed via Procedure \ref{proc} is primitive and has exponent of order $ O(n\log n) $.
\end{theorem}
The proof of Theorem \ref{Thm:primperm} makes use of the following Corollary \ref{cor:fried}, which is a direct consequence of a result of Friedman et al.\  (\cite{Friedman}, Theorem 2.1 and Theorem 2.2). We remind that the \emph{diameter} of a (strongly connected) directed graph $ D=(V,E) $ is equal to $ \max_{u,v\in V} d(u,v) $ where $ d(u,v) $ is the length of the shortest path connecting $ u $ to $ v $.

%of the following group-theoretic result of Friedman et al.\ \cite{Friedman}.

\begin{corollary}[(Friedman et al.\ \cite{Friedman}, Theorem 2.1 and Theorem 2.2)]\label{cor:fried}
Let $ m\!\geq\! 2 $ and $ r\!\geq\! 1 $ be two integers and let $ \lbrace P_1,\dots ,P_m \rbrace$ be a set of $ m $ permutation matrices sampled uniformly and independently at random from $ S_n $.  Let $ D_r$ be the directed graph with vertex set the set of the $ r $-tuples of distinct elements of $ [n] $, having an edge from $ (u_1,u_2,\dots ,u_r ) $ to $ (v_1,v_2,\dots ,v_r ) $ if there exists an $i\in [m]  $ such that $ P_i[u_k,v_k]=1 $ for all $ k=1,\dots ,r $. Then $ D_r $ has diameter of order $ O(\log n) $ with high probability.
%G = (V,E) with vertex set the set of r-tuples of distinct elements of {1,...,n} and there is a directed edge from (u1,u2,···,ur) to (v1,v2,···,vr) iff (v1,v2,···,vr) = (πi(u1),πi(u2),···,πi(ur)) for one of those πi’s. We denote this probability space of random directed graphs Gn,d,r.
%For any $ r\geq 1 $ and $ d\geq 2 $ and for a uniform random sample of $d$ permutation matrices $ P_1,\dots ,P_d $ from $ S_n $, the following property $ F_r $ happens with high probability: for any two $r$-tuples $ v $ and $ w $ of distinct elements in $ [n] $ there is a product of the $ P_i $'s of length $ O(\log n) $ such that $ P(v(j),w(j))>0 $ for all $ j=1,\dots ,r $.% that maps the first $ r $-tuple to the second.
\end{corollary}
Notice that Corollary \ref{cor:fried} also holds in case some of the matrices $ P_i $ are sampled (uniformly) from the set $ \bar{S}_n $.\\
$  $\\
\emph{Proof of Theorem \ref{Thm:primperm}.}\\
It suffices to prove the theorem for $ m=2 $.
Let $ \mathcal{M}\!=\!\lbrace P_1, \bar{P}_2\rbrace $ be a random perturbed permutation set with $\bar{P}_2=P_2+ \mathbb{I}_{i,j}  $ and let $ i'\neq i $ be the integer such that $ P_2[i',j]\!=\!1 $. Corollary \ref{cor:fried} with $ r=2 $ and $ m=2 $ implies that, with high probability, for \emph{any} indices $ v_1,v_2,w_1,w_2\in [n] $ there exists a product $ Q $ of elements of $ \mathcal{M} $ of length $ O(\log n) $ such that $ Q[v_1,w_1]>0 $ and $ Q[v_2,w_2]>0 $; we call this property $ F_2 $.  
We now construct a product of elements of $ \mathcal{M} $ whose $ j $-th column is entrywise positive; to do so we proceed recursively by constructing at each step a product that has one more positive entry in the $ j $-th column than in the previous step. We will then construct from it a positive product.\\
The matrix $ \bar{P}_2 $ has two ones in its $ j $-th column; let $ a_1 $ and $ b_1 $ be two indices such that $ \bar{P}_2[a_1,j]=0 $ and $ \bar{P}_2[a_1,b_1]=1 $ (they do exist as the matrices are NZ). By property $ F_2 $ there exists a product $ Q_1 $ of elements in $ \mathcal{M} $ such that $Q_1[j,i]>0$ and $ Q_1[b_1,i']>0 $; then the product $ \bar{P}_2Q_1\bar{P}_2:=K_1 $ has at least three positive entries in its $ j $-th column. Let now $ a_2 $ and $ b_2 $ be two indices such that $ K_1[a_2,j]\!=\!0 $ and $ K_1[a_2,b_2]>0 $; by property $ F_2 $ there exists a product $ Q_2 $ such that $ Q_2[j,i]>0 $ and $ Q_2[b_2,i']>0 $ and so the product $ K_1Q_2\bar{P}_2:=K_2  $ has at least four positive entries in its $ j $-th column. 
By iterating this procedure, it is clear that $ K_{n-2} $ has a positive column in position $ j $. As each product $ Q_i$ has length $ O(\log n) $, $ K_{n-2} $ has length $ O(n\log n) $. The same reasoning can be applied to the set $ \mathcal{M}^T =\lbrace P^T_1, \bar{P}^T_2\rbrace $ since it is still a perturbed permutation set with $ \bar{P}^T_2=P^T_2+ \mathbb{I}_{j,i}$: there exist products $ T_1,T_2,\dots ,T_{n-2} $ of elements in $ \mathcal{M}^T $ of length $ O(\log n) $ such that, by setting $ W_1\!=\!\bar{P}^T_2T_1\bar{P}^T_2 $ and $ W_s\!=\!W_{s-1}T_s\bar{P}^T_2 $ for $ s=2,\dots ,n-2 $, the final product $ W_{n-2} $ has length $ O(n\log n) $ and its $ i $-th column is entrywise positive. Finally, by property $ F_2 $ there exists a product $ S $ of elements in $ \mathcal{M} $ of length $ O(\log n) $ such that $ S[j,i]>0 $. Then $K_{n-2} SW^T_{n-2} $ is a positive product of elements in $ \mathcal{M} $ of length $ O(n\log n) $.$\quad\quad\,\square$
%Therefore,
%\begin{equation}\label{eq:primhp}
%\mathbb{P}(\mathcal{M}\text{ is primitive})\geq \mathbb{P}\bigl(\exists \text{ products } Q_1,\dots, Q_{n-1},T_1,\dots ,T_{n-2},S\text{ of length }O(\log n)\bigr)
%\end{equation}
%where the righ-hand side of (\ref{eq:primhp}) tends to $ 1 $ as $ n\rightarrow\infty $ by Theorem \ref{thm:perm}.
$  $\\
Theorem \ref{Thm:primperm} shows that Procedure \ref{proc} will (almost surely) never lead to automata with large reset threshold (see Definition \ref{def:assoc_autom} and Theorem \ref{thm:autom_matrix}). In the following section we present a similar result for \emph{random binary sets}.
\subsection{Random sets of binary matrices}\label{subsec:randbin}
We here rephrase some standard notions used in random graph theory (see \cite{randgraphs}, Section 1.5-1.6) in terms of sets of binary matrices; we refer the reader to \cite{randgraphs} for a detailed review on random graphs. Given a property $ \mathscr{P} $ and a set $\mathcal{B}=\lbrace B_1,\dots ,B_m\rbrace$ of binary matrices,  we write $\mathcal{B}\in \mathscr{P}  $ to indicate that the set $ \mathcal{B} $ has the property $ \mathscr{P} $. A property $ \mathscr{P} $ is said to be \emph{increasing} if for any matrix sets $ \mathcal{B}=\lbrace B_1,\dots ,B_m\rbrace $ and $ \mathcal{B}'=\lbrace B'_1,\dots ,B'_m\rbrace $ such that $ \forall\, i=1,\dots ,m $, $ B'_i\leq B_i$, $\mathcal{B}'\in \mathscr{P}  $ implies $\mathcal{B}\in \mathscr{P}  $.\\
%$\mathcal{B}\in \mathscr{P}  $ whenever there exists a set $ \mathcal{B}'=\lbrace B'_1,\dots ,B'_m\rbrace$ such that $ \forall i=1,\dots ,m $, $ B'_i\leq B_i$ and $\mathcal{B}'\in \mathscr{P}  $.\\ %Clearly, the property of being primitive is an increasing property. \\
%Let $ p:\mathbb{N}\rightarrow [0,1] $ be a sequence of numbers between $ 0$ and $ 1 $.
We denote with $ B(n,p) $ an $ n\times n $ random binary matrix where each entry is independently set to $ 1 $ with probability $ p $ and to $ 0 $ with probability $ 1-p $; we denote with $ \mathcal{B}_m(n,p)=\lbrace B_1(n,p),\dots ,B_m(n,p)\rbrace $ a set of $ m\geq 2 $ matrices obtained independently in this way. The parameter $ p $ may depend on the matrix size $ n $, so it has to be intended as a sequence of real numbers $ p(n)\in[0,1] $, $ n\in\mathbb{N} $; to ease the notation, we will sometimes avoid to explicit the dependancy of $ p $ on $ n $, so we will write $ B(n,p) $ instead of $ B(n,p(n)) $ and $ \mathcal{B}_m(n,p)$ instead of $ \mathcal{B}_m(n,p(n))$.
\begin{definition}
Given an increasing property $ \mathscr{P} $, a sequence $ \hat{p}(n)\in [0,1]$, $ n\in\mathbb{N} $, is called a \emph{threshold} for the random binary set $ \mathcal{B}_m(n,p) $ \emph{with respect to} $ \mathscr{P} $ if, for any sequence $ p(n)\in [0,1]$, $ n\in\mathbb{N} $:
\begin{equation*}
\lim_{n\rightarrow\infty}\mathbb{P}\Bigl(\mathcal{B}_m\bigl(n,p(n)\bigr)\in  \mathscr{P}\Bigr )=\begin{cases}
1 & \text{ if } p\gg \hat{p},\\
0 & \text{ if } p\ll \hat{p}, 
\end{cases}
\end{equation*}
where $ p\gg \hat{p} $ if and only if $ \lim_{n\rightarrow\infty}\hat{p}(n)/p(n)=0 $. Furthermore, a sequence $ \hat{p}(n)\in [0,1]$, $ n\in\mathbb{N} $,  is said to be a \emph{sharp threshold} for the random binary set $ \mathcal{B}_m(n,p) $ \emph{with respect to} $ \mathscr{P} $ if for any sequence $ p(n)\in [0,1]$, $ n\in\mathbb{N} $, and for every fixed $ \alpha >0 $: 
\begin{equation*}
\lim_{n\rightarrow\infty}\mathbb{P}\Bigl(\mathcal{B}_m\bigl(n,p(n)\bigr)\in  \mathscr{P} \Bigr)=\begin{cases}
1 & \text{ if } \,\exists N\!\in\!\mathbb{N}:\,\forall n>N,\, p(n) \geq (1+\alpha )\hat{p}(n)\\
0 & \text{ if } \,\exists N\!\in\!\mathbb{N}:\,\forall n>N,\, p(n) \leq (1-\alpha )\hat{p}(n).
\end{cases}
\end{equation*}
\end{definition}
A (sharp) threshold thus represents a \emph{phase transition} for $ \mathcal{B}_m(n,p) $ from not having property $ \mathscr{P} $ with high probability to having property $ \mathscr{P} $ with high probability.
\begin{remark}
Note that thresholds are in general defined up to the asymptotic relation $ \hat{p}'=\Theta(\hat{p}) $; in other words, if $ \hat{p} $ is a threshold, than so is every sequence $ \hat{p}'(n)\in [0,1]$, $ n\in\mathbb{N} $, for which there exist $ C,c>0 $ and $ N\in\mathbb{N} $ such that $\forall\, n\geq N $, $ c\hat{p}'(n)\leq \hat{p}(n) \leq C\hat{p}'(n) $. This implies that a threshold is never uniquely defined, despite it is customary to call it \emph{the} threshold (see for example \cite{bollobas}, \cite{randgraphs}). %, it should be kept in mind that it is defined only within constant factors. 
The same can be said about a \emph{sharp} threshold $ \hat{p} $: in this case, any sequence $ \hat{p}'(n)\in [0,1]$, $ n\in\mathbb{N} $, such that $ \lim_{n\rightarrow\infty} \hat{p}'(n)/\hat{p}(n)=1 $ is as well a sharp threshold.
\end{remark}
We denote with $\mathcal{PR}$ the property for a binary matrix set to be \emph{primitive}; it is easy to prove that it is an increasing property. The following theorem establishes a sharp threshold for $\mathcal{B}_m(n, p)$ to be primitive and provides an asymptotic estimate of the expected exponent of $\mathcal{B}_m(n, p)$ when it is a primitive NZ-set.
\begin{theorem}\label{thm:binarymat}
Let $ m\geq 2 $ be an integer, $ c\in\mathbb{R} $ and $ \hat{p}(n)=(\log n+c)/n $. Then the sequence $ \hat{p} $ is a \emph{sharp threshold} for $ \mathcal{B}_m(n, p)$ with respect to $ \mathcal{PR} $. Moreover,
\begin{equation}\label{eq:a(m,c)}
 a(m,c)\leq\lim_{n\rightarrow\infty} \mathbb{P}\Bigl(\mathcal{B}_m\bigl(n, \hat{p}(n)\bigr)\in  \mathcal{PR} \Bigr)\leq 1-\bigl( 1-e^{-e^{-c}}\bigr)^m,
\end{equation}
where $ a(m,c)=1-\bigl( 1-e^{-2e^{-c}}\bigr)^m-me^{-2e^{-c}}\bigl( 1-e^{-2e^{-c}}\bigr)^{m-1} $. \\In addition:
\begin{enumerate}
\item If $ p(n)\in [0,1] $, $ n\in\mathbb{N} $, is such that $ \exists\, \alpha \!>\!0, N\!\in\!\mathbb{N}: \forall\, n\!>\!N,\, p(n)\!\geq\! (1+\alpha)\hat{p}(n)$, then $exp\bigl(\mathcal{B}_m(n, p)\bigr)= O(n\log n) $ with high probability;
\item $exp\bigl(\mathcal{B}_m(n,\hat{p})\bigr)= O(n\log^3 n) $ with high probability, under the condition that $\mathcal{B}_m(n, \hat{p})$ is an NZ-primitive set. %; in other words as $n\rightarrow +\infty  $,
\end{enumerate}
\end{theorem}
Note that Theorem \ref{thm:binarymat} implies that for any constant sequence $ p(n)\equiv q\in [0,1] $, the set $ \mathcal{B}_m(n,q) $ has a positive product of length $ O(n\log n) $ with high probability; in particular, $ q=1/2 $ induces the uniform distribution over the set of the binary matrix sets of cardinality $ m $.
Before proving Theorem \ref{thm:binarymat} we need two preliminary results, the following Lemma \ref{lem:unifdistr} and Theorem \ref{thm:nicaud}, the latter presented by Nicaud in \cite{Nicaud}.
\begin{lemma}\label{lem:unifdistr}
Let $ \mathscr{C} $ be a finite set of $ n\!\times\! n $ binary matrices such that one of the following properties hold:
\begin{enumerate}
\item for all $ P,Q\in S_n $, $  \mathscr{C}=\lbrace PCQ: C\in\mathscr{C}\rbrace := P\mathscr{C}Q $ and for all $ \,C,D\in \mathscr{C}$, there exist $\, T_1,T_2\in S_n $ such that $ C=T_1DT_2 $ ;
\item $ \mathscr{C}=\mathcal{R}_n $ ;
\item $ \mathscr{C}=\mathcal{C}_n $ .
\end{enumerate}
Let $ X_{\mathscr{C}} $ be a random variable with values in $ \mathscr{C}\cup \lbrace 0\rbrace $, defined in the following way: a random binary matrix $ B(n,p) $ is generated, then $ X\!=\!0 $ if $ B(n,p) $ does not dominate any matrix in $ \mathscr{C} $, otherwise $ X\!=\!C $ with $ C $ sampled uniformly among the elements of $ \mathscr{C} $ dominated by $ B(n,p) $. Let $ \mathbb{P}_{X_{\!\mathscr{C}}} $ be the distribution of $ X_\mathscr{C} $. Then it holds that, for any $ C,D\in \mathscr{C} $:
\begin{equation}\label{eq:unifdistr}
\mathbb{P}_{X_{\!\mathscr{C}}}(C)=\mathbb{P}_{X_{\!\mathscr{C}}}(D).
\end{equation}

\end{lemma}
\begin{proof}
Suppose first that $($\scriptsize $\mathrm{I} $\normalsize$)$ holds. %For any $ T_1,T_2\in S_n $, the function $ f_{T_1,T_2}\!: \mathscr{C}\rightarrow \mathscr{C} $, $f_{T_1,T_2}(C)= T_1CT_2 $, is injective since $ T_1$ and $T_2$ are invertible matrices. As $ \mathscr{C} $ is finite, $ f_{T_1,T_2}$ is bijective. This implies that for any $ T_1,T_2\in S_n $, 
%\begin{equation}\label{eq:randvar}
% \mathscr{C}=\lbrace T_1CT_2: C\in\mathscr{C}\rbrace := T_1\mathscr{C}T_2.
%\end{equation}
Let $ \mathbb{P} $ be the distribution of $ B(n,p) $; we write $ \mathbb{P}(M) $ for $ \mathbb{P}\bigl(B(n,p)\!=\!M\bigr) $. By definition, for any $ C\!\in\! \mathscr{C} $, $\mathbb{P}_{X_{\!\mathscr{C}}}(C)=\sum_{M\geq C}\mathbb{P}(M) \vert \lbrace C'\!\in\! \mathscr{C}: M\!\geq\! C'\rbrace\vert^{-1}  $, where $ M $ is taken in the set of the binary matrices. Let $ C,D\!\in\! \mathscr{C}  $ and $ T_1,T_2\in S_n  $ such that $ C=T_1DT_2 $. Observe that $ \mathbb{P}(M) $ depends only on the number of positive entries of $ M $ so $ \mathbb{P}(M)=\mathbb{P}(T_1^{-1}MT_2^{-1}) $ as $ T_1 $ and $ T_2 $ are permutations. It follows that
\begin{align*}
\mathbb{P}_{X_{\!\mathscr{C}}}(C)&=\sum_{M\geq T_1DT_2}\,\mathbb{P}(T_1^{-1}MT_2^{-1})\, \vert \lbrace C'\!\in\! \mathscr{C}: M\geq C'\rbrace\vert^{-1}=\\
&=\sum_{T_1^{-1}MT_2^{-1}\geq D}\,\mathbb{P}(T_1^{-1}MT_2^{-1})\, \vert \lbrace C'\!\in\! T_1^{-1}\mathscr{C}T_2^{-1}: T_1^{-1}MT_2^{-1}\geq C'\rbrace\vert^{-1}=\\
&=\mathbb{P}_{X_{\!\mathscr{C}}}(D).
\end{align*}
Suppose now that $($\scriptsize $\mathrm{II} $\normalsize$)$ holds. We show that $ \mathbb{P}_{X_\mathscr{C}}(C) $ does not depend on $ C\in\mathscr{C}$ and so (\ref{eq:unifdistr}) must hold. Let $ a=(a_1,\dots ,a_n) $ be a vector in $ [n]^n $; we write that $ M=a $ if the $ i $-th row of $ M $ has exactly $ a_i $ positive entries. Then,
\begin{align*}
\mathbb{P}_{X_\mathscr{C}}(C)&=\sum_{a_1=1}^n \cdots \sum_{a_n=1}^n\sum_{\substack{M=a,\\ M\geq C}} \dfrac{\mathbb{P}(M)}{\vert \lbrace C': M\geq C'\rbrace\vert}\\
&=\sum_{a_1=1}^n \cdots \sum_{a_n=1}^n \prod_{i=1}^n a_i^{-1} \binom{n-1}{a_i-1} p^{a_i}(1-p)^{n-a_i}.
\end{align*}
%2^{-n^2}\sum_{a_1=1}^n \cdots \sum_{a_n=1}^n \sum_{\substack{M=a,\\ M\geq R}} \prod_{i=1}^n a_i^{-1}=\\ 
Case $($\scriptsize $\mathrm{III} $\normalsize$)$ can be proved analogously. 
\end{proof}
\begin{theorem}[(\cite{Nicaud}, Theorem 3)]\label{thm:nicaud}
Let $ \mathcal{A} $ be a random $ n $-state DFA of $ m\geq 2 $ letters where each letter is chosen independently and uniformly at random from $ \mathcal{R}_n $. Then $ \mathcal{A} $ admits a synchronizing word of length $ O(n\log^3 n) $ with high probability.
\end{theorem}
\textit{Proof of Theorem \ref{thm:binarymat}.}\\
With a slight abuse of notation we denote with $ \mathbb{P} $ the distribution of $ \mathcal{B}_m(n, p) $.\\
Suppose first that there exists $\alpha \!>\!0$ and $N\!\in\!\mathbb{N}$ such that $ \forall\, n>N,\, p(n)\geq (1+\alpha)\hat{p}(n)$. We need to prove that 
\[ \lim_{n\rightarrow\infty}\mathbb{P}\biggl(\mathcal{B}_m\bigl(n, p(n)\bigr)\!\in\!\mathcal{PR}\, \text{ and } \,exp\Bigl(\mathcal{B}_m\bigl(n, p(n)\bigr)\Bigr)\!=\!O(n\log n) \biggr)=1. \]
Without loss of generality we can just consider the case $ m=2$. We first show that $ B(n,p) $ dominates a permutation matrix with high probability and it also dominates a perturbed permutation matrix\footnote{We remind that a \emph{perturbed permutation matrix} is a permutation matrix where one of its $ 0 $-entries has been changed into a $ 1 $.} with high probability. This will imply that the random set $\mathcal{B}_2(n, p)  $ dominates\footnote{We say that a matrix set $ \mathcal{M}=\lbrace M_1,\dots ,M_m\rbrace $ dominates a matrix set $ \mathcal{M}'=\lbrace M'_1,\dots ,M'_m\rbrace $ if for all $ i=1,\dots ,m $, $ M_i\geq M'_i $.} a perturbed permutation set with high probability.
Let $ G(n,n,p) $ be a random bipartite graph with vertex set $ V=V_1\cup V_2 $ where $V_1\!=\![n]\!=\!V_2   $, and edge set $ E $ such that $ (i,j)\in E $ if and only if $ i\in V_1 $, $ j\in V_2 $ and $ B(n,p)[i,j]=1 $. Equivalently, in the bipartite graph $ G(n,n,p) $ there is an edge between vertices $ i\in V_1 $ and $ j\in V_2 $ with probability $ p(n) $. A \emph{perfect matching} of a graph is a subset $ E'$ of its egdes such that exactly one edge in $ E' $ is incident to each vertex of the graph: it is easy to see that $ G(n,n,p) $ admits a perfect matching if and only if $ B(n,p) $ dominates a permutation matrix. Theorem 4.1 in \cite{randgraphs} shows that, under our hypothesis, $ G(n,n,p) $ admits a perfect matching with high probability; consequently, $ B(n,p) $ dominates a permutation matrix with high probability. It is now easy to prove that $ B(n,p) $ dominates a perturbed permutation matrix with high probability: indeed this probability is equal to the probability that $ B(n,p) $ dominates a permutation matrix minus the probability that $ B(n,p) $ \textit{is} a permutation matrix. The former term goes to $ 1 $ as $ n $ tends to infinity as proved above while the latter term is smaller than $ \bigl( np(n)\bigl(1-p(n)\bigr)^{n-1}\bigl) ^n  $, which tends to $ 0 $ as $ n $ goes to infinity.\\
We now use Theorem \ref{Thm:primperm} on the perturbed permutation set dominated by $ \mathcal{B}_2(n, p) $.
Both the sets $ S_n $ and $ \bar{S}_n  $ satisfy the hypothesis $($\scriptsize $\mathrm{I} $\normalsize$)$ of Lemma \ref{lem:unifdistr}; let $ X=X_{S_n} $ and $ \bar{X}=X_{\bar{S}_n} $ be the random variables defined in the same lemma. We assume $ X $ and $ \bar{X} $ to be independent. Lemma \ref{lem:unifdistr} implies that for every $ P\!\in\! S_n  $, $
\mathbb{P}_X(P)=\bigl(1-\mathbb{P}_X( 0)\bigr)/ n!$ and for every $ \bar{P}\!\in\! \bar{S}_n  $,
%\begin{equation}\label{eq:almostunif}
$ \mathbb{P}_{\bar{X}}(\bar{P})=\bigl(1-\mathbb{P}_{\bar{X}}( 0)\bigr)/(n!\,n(n-1))
$; indeed one can verify that $ \vert \bar{S}_n\vert= n!\,n(n-1)$. The fact that $ B(n,p) $ dominates a permutation matrix with high probability implies that $ \mathbb{P}_X(0)\longrightarrow 0 $ as $ n\rightarrow +\infty $ and the fact that $ B(n,p) $ dominates a perturbed permutation matrix with high probability implies that $ \mathbb{P}_{\bar{X}}(0)\longrightarrow 0 $ as $ n\rightarrow +\infty $. Let $ \mathbb{P}_{X \times \bar{X}}= \mathbb{P}_X\cdot \mathbb{P}_{\bar{X}}$ be the joint distribution of $ X $ and $ \bar{X} $ on $ S_n\times \bar{S}_n $ and let $ \Omega\subset S_n\times \bar{S}_n $ be the event that a perturbed permutation set of cardinality $ 2 $ is primitive and with exponent of order $ O(n\log n) $. Since $\mathcal{PR}  $ is an increasing property, it holds that:
\begin{equation}\label{eq:upperb}
\mathbb{P}\Bigl(\mathcal{B}_m(n, p)\in  \mathcal{PR} \,\text{ and }\, exp\bigl(\mathcal{B}_m(n, p)\bigr)=O(n\log n) \Bigr)\geq 
\mathbb{P}_{X\times\bar{X}}( \Omega )
\end{equation}
and
\begin{equation}\label{eq:binmat}
\mathbb{P}_{X\times\bar{X}}( \Omega )= \bigl(1-\mathbb{P}_{X}( 0)\bigr)\bigl(1-\mathbb{P}_{\bar{X}}( 0)\bigr)\!\!\!\sum_{\lbrace P_1,\bar{P}_2\rbrace\in \Omega  }\!\!\! (n!)^{-1} \bigl(n!\, n(n-1)\bigr)^{-1}.
\end{equation}
The summation in the right-hand side of (\ref{eq:binmat}) is the probability that a set of cardinality $ 2 $ generated by Procedure \ref{proc} is primitive and with exponent of order $ O(n\log n) $, which goes asymptotically to 1 by Theorem \ref{Thm:primperm}. Since $ \mathbb{P}_{X}(0) $ and $ \mathbb{P}_{\bar{X}}(0) $ tend to zero as $ n $ goes to infinity, eq. (\ref{eq:binmat}) goes asymptotically to $ 1 $. In view of the inequality (\ref{eq:upperb}), we conclude.\\
Suppose now that there exist $\alpha \!>\!0$ and $N\!\in\!\mathbb{N}$ such that $ \forall\, n>N,\, p(n)\leq (1-\alpha)\hat{p}(n)$. We need to prove that $ \lim_{n\rightarrow\infty}\mathbb{P}\bigl(\mathcal{B}_m(n, p(n))\!\in\!\mathcal{PR}\bigr)=0.$
If every matrix of a set has a zero-row, the set cannot be primitive: we show that $ \mathcal{B}_m(n, p) $ has this property with high probability. Indeed, this probability is equal to $(1-(1-(1-p(n))^n)^n)^m $; by hypothesis $ (1-p(n))^n\rightarrow 0 $ as $ n\rightarrow\infty $, so
$
(1-(1-p(n))^n)^n \sim e^{-ne^{-p(n)n}}
$ that tends asymptotically to $ 0 $.\\
It remains to prove (\ref{eq:a(m,c)}) and $($\scriptsize $\mathrm{II} $\normalsize$)$; we start by proving (\ref{eq:a(m,c)}). The term $ 1-\mathbb{P}\bigl(\mathcal{B}_m(n, \hat{p})\!\in\!  \mathcal{PR} \bigr)=\mathbb{P}\bigl(\mathcal{B}_m(n, \hat{p})\!\notin\!  \mathcal{PR} \bigr) $ is lower bounded by the probability that each matrix in $\mathcal{B}_m(n, \hat{p})  $ has at least a zero row, which is equal to $ \bigl( 1\!- \mathbb{P}\bigl(B(n,\hat{p})\text{ has no zero rows}\bigr)\bigr)^m $. The probability that $ B(n,\hat{p}) $ has exactly $ k $ zero-rows is a binomial distribution of parameters $ n $ and $ q(n)=(1-\hat{p}(n))^n $, which converges to a Poisson distribution of mean $ \mu= e^{-c}  =\lim_{n\rightarrow\infty} nq(n) $. This implies that $ \mathbb{P}(B(n,\hat{p})\text{ has no zero rows})$ converges asymptotically to $ e^{-e^{-c}} $, and so $1-\lim_{n\rightarrow\infty}\mathbb{P}\bigl(\mathcal{B}_m(n, \hat{p}(n))\!\in\!  \mathcal{PR} \bigr)\geq \bigl( 1-e^{-e^{-c}}\bigr)^m$ which proves the upper bound in (\ref{eq:a(m,c)}). For the lower bound, let $ D $ be the event that there exist at least two matrices in $ \mathcal{B}_m(n, \hat{p}) $ such that each of them dominates a permutation matrix; it holds that $ \mathbb{P}\bigl(\mathcal{B}_m(n, \hat{p})\in  \mathcal{PR}\bigr)\!\geq\! \mathbb{P}\bigl(\mathcal{B}_m(n, \hat{p})\!\in\!  \mathcal{PR}\,\vert\, D\bigr)\mathbb{P}(D)  $. The term $ \mathbb{P}(\mathcal{B}_m(n, \hat{p})\!\in\!  \mathcal{PR}\,\vert\, D) $ tends asymptotically to $ 1 $: this can be proved similarly as in the case where $ p\geq (1+\alpha)\hat{p} $, by introducing the random variables $ X=X_{S_n} $ and $ \bar{X}=X_{\bar{S}_n} $ as in Lemma \ref{lem:unifdistr}. The difference is that now $ B(n,\hat{p}) $ is generated conditioned to the fact that it dominates a permutation matrix so $ X $ takes value in $ S_n $; eq.(\ref{eq:unifdistr}) still holds and so we can apply Theorem \ref{Thm:primperm}. It then remains to show that $ \mathbb{P}(D) $ tends to $ a(m,c) $ as $ n\rightarrow\infty $; this is straightforward as the probability that $ B(n,\hat{p}) $ dominates a permutation matrix tends asymptotically to $ e^{-2e^{-c}} $ (\cite{randgraphs}, Theorem 4.1).\\
Finally, we prove item $($\scriptsize $\mathrm{II} $\normalsize$)$. We can suppose $ m=2 $ without loss of generality. Let $ X_r=X_{\mathcal{R}_n} $ and $ X_c=X_{\mathcal{C}_n} $ be the random variables defined in Lemma \ref{lem:unifdistr}. By hypothesis the sampled set $ \mathcal{B}_2(n,\hat{p}) $ is known to be NZ, so Lemma \ref{lem:unifdistr} implies that $ \mathbb{P}_{X_r} $ is the uniform distribution over $ \mathcal{R}_n $ and $ \mathbb{P}_{X_c} $ is the uniform distribution over $ \mathcal{C}_n $. By Theorem \ref{thm:nicaud}, we have that with high probability $ \mathcal{B}_2(n,\hat{p}) $ admits  a product $ C $ of length $ O(n\log^3 n) $ with a positive column (say in position $ j $) and with high probability $ \mathcal{B}_2(n,\hat{p}) $ admits a product $ R $ of length $ O(n\log^3 n) $ with a positive row (say in position $ i$). 
 %The same reasoning can be applied to $ \mathcal{B}^T_2(p) $ as it is NZ as well; with high probability it admits a product of length $ O(n\log^3 n) $ with a positive column, which means that  $ \mathcal{B}_2(p) $ has a product of length $ O(n\log^3 n) $ with a positive row. 
Since $ \mathcal{B}_2(n,\hat{p}) $ is primitive by hypothesis, its underlying graph is strongly connected, which means that there exists a product $ L $ of elements of $ \mathcal{B}_2(n,\hat{p}) $ of length at most $ n-1 $ such that $ L[j,i]>0 $. The product $ CLR $ is a positive product of length $ O(n\log^3 n) $ so $($\scriptsize $\mathrm{II} $\normalsize$)$ follows. % $ \mathcal{B}_2(n,\hat{p}) $ admits a positive product of length $ O(n\log^3 n) $ with high probability.
$\square$\\
$  $\\
Notice that, since primitivity is not influenced by the actual values of the positive entries of the matrices, Theorem \ref{thm:binarymat} is naturally extended to random nonnegative matrices. Furthermore, in the case $p=\hat{p} $, both the left-hand term and the right-hand term of eq.(\ref{eq:a(m,c)}) approaches $ 1 $ as the number of matrices $ m $ increases, which is reasonable to expect. We underline that the difference in the upper bounds on $ exp\bigl( \mathcal{B}_m(n, p)\bigr) $ that we get when $ p=\hat{p} $ or when $ p\geq (1+\alpha)\hat{p} $ is due to the fact that it is not possible to use the same reasoning. Indeed, when $ p=\hat{p} $ the probability that $ B(n,\hat{p}) $ dominates a permutation matrix is asymptotically equal to a constant strictly smaller than $ 1 $ (\cite{randgraphs}, Theorem 4.1) and so we cannot make use of Theorem \ref{Thm:primperm} anymore. Notice also that the condition that all the matrices of the set are NZ is weaker than requiring that all the matrices of the set dominate a permutation matrix: it is indeed easy to build an NZ matrix that does not dominate a permutation matrix.\\
%that is not true that any NZ-matrix dominates a permutation matrix.\\
It is interesting to compare Theorem \ref{thm:binarymat} with a result of Gerencs\'{e}r et al.\ (\cite{GerenGusJung}, Corollary 3); they prove that $ \lim_{n\rightarrow\infty}\log (exp(n))/n=(\log 3)/3 $ where $exp(n)\!=\!\max \lbrace exp(\mathcal{M}): \mathcal{M} \text{ is a primitive set of }n\!\times\! n\text{ matrices}\rbrace $. Our result shows that the sets whose exponent reaches $ \exp(n) $ must be very few and that they are almost impossible to be attained by $ \mathcal{B}_m(n, p) $ and in particular from a uniform distribution; indeed the \emph{average} exponent is much smaller. % when $ p>(\log n+c)/n$ asymptotically.
\\Summarizing, in view of the connection between primitive sets and synchronizing DFAs established by Theorem \ref{thm:autom_matrix}, Theorem \ref{thm:binarymat} suggests that there is very little hope of generating slowly synchronizing automata from $ \mathcal{B}_m(n, p) $, no matter how the sequence $ p(n) $ behaves.
\subsection{Random NDFAs}\label{subsec:randNDFA}
The random binary set $ \mathcal{B}_m(n,p) $ can be seen as a random nondeterministic finite automaton. We here apply Theorem \ref{thm:binarymat} to the $ 2 $-directability and $ 3 $-directability properties of $ \mathcal{B}_m(n,p) $.

\begin{corollary}
Let $ m\geq 2 $ an integer and $ \hat{p}(n)=(\log n+c)/n $ for some $ c\in\mathbb{R} $. Let $ p(n)\in [0,1]$, $ n\in\mathbb{N} $, be a sequence such that there exist $ \alpha \!>\!0$ and $ N\!\in\!\mathbb{N}:\, \forall\, n\!>\!N,\, p(n)\!\geq\! (1+\alpha)\hat{p}(n)$. Then with high probability $\, \mathcal{B}_m(n,p) $ is $ 2 $-directable and $ d_2\bigl( \mathcal{B}_m(n,p)\bigr)=O(n\log n) $. In particular, for any fixed integer $ m\geq 2 $, with high probability an $ m $-letter NDFA generated according to the uniform distribution is $ 2 $-directable and has a $ 2 $-directing word of length $ O(n\log n) $.
\end{corollary}

\begin{proof}
It is a straightforward consequence of Therorem \ref{thm:binarymat} and (\ref{eq:d1d2}). The uniform distribution is obtained by choosing $ p(n)=1/2 $ for every $ n\in\mathbb{N} $.
\end{proof}
The following corollary shows that $ \hat{p}(n)=(\log n+c)/n $ is as well a sharp threshold for $ \mathcal{B}_m(n,p)$ with respect to the $ 3 $-directability property:

\begin{corollary} 
Let $ m \geq 2 $ be an integer, $ c\in\mathbb{R} $ and $ \hat{p}(n)=(\log n+c)/n $. The sequence $ \hat{p} $ is a \emph{sharp threshold} for $ \mathcal{B}_m(n,p) $ with respect to the $ 3 $-directability property. It also holds that
\begin{equation}\label{eq:scrambl}
a(m,c)\leq\lim_{n\rightarrow\infty} \mathbb{P}\Bigl(\mathcal{B}_m\bigl(n, \hat{p}(n)\bigr)\text{ is $ 3 $-directable} \Bigr)\leq 1-\bigl( 1-e^{-e^{-c}}\bigr)^m.
\end{equation}
where $ a(m,c)= 1-\bigl( 1-e^{-2e^{-c}}\bigr)^m-me^{-2e^{-c}}\bigl( 1-e^{-2e^{-c}}\bigr)^{m-1} $. \\Furthermore:
\begin{enumerate}
\item If $ p(n)\in [0,1] $, $n\in\mathbb{N}  $, is such that $ \exists\, \alpha \!>\!0, N\!\in\!\mathbb{N}: \forall\, n\!>\!N,\, p(n)\!\geq\! (1+\alpha)\hat{p}(n)$, then $d_3\bigl(\mathcal{B}_m(n, p)\bigr)= O(n\log n) $ with high probability;
\item $d_3\bigl(\mathcal{B}_m(n,\hat{p})\bigr)= O(n\log^3 n) $ with high probability, under the condition that $\mathcal{B}_m(n, \hat{p})$ is an NZ-primitive set. %; in other words as $n\rightarrow +\infty  $,
\end{enumerate}
In particular, for any fixed integer $ m\geq 2 $, with high probability an $ m $-letter NDFA generated according to the uniform distribution is $ 3 $-directable and has a $ 3 $-directing word of length $ O(n\log n) $.
\end{corollary}

\begin{proof}
If the sequence $ p(n) $ is such that there exist $ \alpha \!>\!0$ and $ N\!\in\!\mathbb{N}$ such that $ \forall\, n\!>\!N,\, p(n)\!\geq\! (1+\alpha)\hat{p}(n)$, then by Theorem \ref{thm:binarymat} and (\ref{eq:d1d2}) it holds that \[ \lim_{n\rightarrow\infty}\mathbb{P}\Bigl(\mathcal{B}_m\bigl(n,p(n)\bigr)\text{ is $ 3 $-directable}\,\text{ and } d_3\bigl(\mathcal{B}_m\bigl(n, p(n)\bigr)\bigr)= O(n\log n) \Bigr)=1. \]
If there exist $ \alpha \!>\!0$ and $ N\!\in\!\mathbb{N}$ such that $ \forall\, n\!>\!N,\, p(n)\!\leq\! (1-\alpha)\hat{p}(n)$, then $ \lim\limits_{n\rightarrow\infty}\mathbb{P}\bigl(\mathcal{B}_m(n,p(n))\text{ is $ 3 $-directable}\bigr)\!=\! 0 $ due to the same argument used in the proof of Theorem \ref{thm:binarymat}: with high probability all the matrices of $\mathcal{B}_m(n,p)  $ have a zero-row.\\
Theorem \ref{thm:binarymat} also trivially implies the lower bound in (\ref{eq:scrambl}) and item $($\scriptsize $\mathrm{II} $\normalsize$)$, since a positive product has in particular an entrywise positive column.\\
It remains to prove the upper bound in (\ref{eq:scrambl}). In the proof of Theorem \ref{thm:binarymat} we have seen that the asymptotic probability for $ \mathcal{B}_m(n, \hat{p}) $ to have each matrix with a zero-row is equal to $ ( 1\!-\!e^{-e^{-c}})^m $, in which case $ \mathcal{B}_m(n, \hat{p}) $ is not $ 3 $-directable. Therefore, $ \lim_{n\rightarrow\infty} \mathbb{P}(\mathcal{B}_m(n, \hat{p}(n))\text{ is $ 3 $-directable})\!\leq 1\!-\!( 1\!-\!e^{-e^{-c}})^m $.\\
The uniform distribution is obtained by choosing $ p(n)=1/2 $ for every $ n\in\mathbb{N} $.
%Case $ p\!=\!\hat{p} $: primitivity implies the existence of a scrambling product so 
%$\lim\limits_{n\rightarrow\infty} \mathbb{P}(\mathcal{B}_m(n, p)\!\in\!  \mathcal{PR} )\leq \lim\limits_{n\rightarrow\infty} \mathbb{P}(\mathcal{B}_m(n, p)\!\in\!  \mathcal{SC} )$. The lower bound in (\ref{eq:scrambl}) is hence proven by applying Theorem \ref{thm:binarymat}.
\end{proof}
Notice again that, for any fixed $ c\in\mathbb{R} $, the right-hand term and the left-hand term of (\ref{eq:scrambl}) both tend to $ 1 $ as the number of matrices $ m $ (the cardinality of the alphabet of the NDFA) increases.

\section{A randomized algorithm for generating proper primitive sets}\label{sec:algorithm}
In this section we describe a randomized procedure to build proper\footnote{We remind that we call a primitive set \emph{proper} if it needs all its matrices to be primitive.} primitive sets making use of the Protasov-Voynov characterization theorem (Theorem \ref{thmProt}, Section \ref{sec:def}), which describes a combinatorial property that an NZ-matrix set must have in order \textit{not} to be primitive: by constructing a primitive set such that each of its proper subsets has this property, we can make it \emph{proper}. In particular, we will build proper perturbed permutation sets, for the reasons presented at the beginning of Section \ref{sec:whp}.\\
Theorem \ref{thmProt} implies that a primitive set of $ m $ matrices is proper if and only if each of its subsets of cardinality $ m-1 $ has a block-permutation structure on a certain partition, so this is the condition we will enforce.
%If we want to find minimally synchronizing automata, Proposition \ref{prop:minsetsautom} tells us that we just need to generate minimally primitive perturbed permutation sets;
%In this section we implement a randomized procedure to build them.\\ % to find minimal primitive sets of (possibly) any cardinality and we show that,
%Theorem \ref{thmProt} says that a matrix set is \textit{not} primitive if all the matrices share the same block-permutation structure, therefore a set of $ m $ matrices is minimally primitive iff every subset of cardinality $ m-1 $ has a block-permutation structure on a certain partition; this is the condition we will enforce. %, by assignign a prtition to each of these subsets. 
As we are dealing with perturbed permutation sets, by Proposition \ref{cor:gonze} these partitions must have blocks of the same size; if the blocks of the partition have size $ n/q $, we call it a \textit{$ q $-partition} and we say that the set has a \textit{$ q $-permutation structure}. The algorithm first generates a set of permutation matrices satisfying the requested block-permutation structures and then a $ 0 $-entry of one of the obtained matrices is changed into a $ 1 $; while doing this last step, we will make sure to preserve all the block-permutation structures of the matrix. We underline that our algorithm finds perturbed permutation sets that, if are primitive, are also proper; the construction itself does not guarantee primitivity and this property has to be verified at the end.\\
One of the advantages of using perturbed permutation sets is that we can easily generate proper synchronizing DFAs from them, as shown by the following proposition:
 %as a set with just permutation matrices will never be primitive, we need to add at least a $ 1 $. 
%Furthermore, Corollary \ref{cor:gonze} will be crucial in the development of the second random construction in Section \ref{sec:2}. \\ 
 %We already explained in Section \ref{sec1} why we are considering perturbed permutation sets; the interest in sets with more than two matrices relies on the fact that in automata theory, to the best of our knowledge, few experiments has been conducting on automata with more than two letters, partially due to the computational cost and mostly because two letters automata are very likely to synchronize (see \cite{Berl}, \cite{Nicaud}) and the lack of a strategy to construct minimal automata. 
%As already anticipated in the introduction, de Bondt et al.\ proved in \cite{BondtDon} that every automaton with $ n\leq 6 $ and any cardinality fulfils the \v{C}ern\'{y} conjecture, but the behaviour of automata with bigger $ n $ is still quite unclear. 
 \begin{proposition}\label{prop:minsetsautom}
Let $ \mathcal{M}=\lbrace P_1, \dots, P_{m-1}, P_m + \mathbb{I}_{i,j}\rbrace $ be a proper primitive perturbed permutation set and let $j'\neq j$ be the integer such that $ P_m[i,j']=1 $. The synchronizing automaton $ \mathcal{A}(\mathcal{M}) $ (see Definition \ref{def:assoc_autom}) can be written as $\mathcal{A}(\mathcal{M})=\lbrace P_1, \dots, P_{m-1}, P_m,M\rbrace$ with $ M\!=\! P_m + \mathbb{I}_{i,j}- \mathbb{I}_{i,j'} $. If $\mathcal{A}(\mathcal{M})  $ is not proper, then $ \bar{\mathcal{A}}=\lbrace P_1, \dots, P_{m-1}, M\rbrace $ is. 
\end{proposition}

\begin{proof}
Suppose $ \mathcal{A}(\mathcal{M}) $ is not proper; the only matrix we can delete from the set without losing synchronization is $ P_m $. Indeed, we cannot delete $ M $ as all the others are permutation matrices. For $ i=1,\dots ,m-1 $, let $ \mathcal{M}_i$ be the set obtained from $ \mathcal{M} $ by erasing $ P_i $; by hypothesis, $ \mathcal{M}_i $ is not primitive so the automaton $ \mathcal{A}(\mathcal{M}_i)$ is not synchronizing. But $ \mathcal{A}(\mathcal{M}_i)$ is indeed the automaton obtained by erasing $ P_i $ from $ \mathcal{A}(\mathcal{M}) $, so $ \bar{\mathcal{A}} $ has to be synchronizing and proper.
\end{proof}
\subsection{The algorithm}
Given $ R,C\!\subset\! [n] $ and a matrix $ M $, we indicate with $ M[R,C] $ the submatrix of $ M $ with rows indexed by $ R $ and columns indexed by $ C $.\\
For generating a set of $ m $ matrices $ \mathcal{M}\!=\!\lbrace M_1,\dots ,M_m\rbrace $ we choose $ m $ prime numbers $ q_1\geq \dots \geq q_m\geq 2 $ and we set $ n\!=\!\prod_{i=1}^m q_i $. For $ j\!=\!1,\dots ,m $, we require the set $ \lbrace M_1,\dots ,M_{j-1}, M_{j+1},\dots ,M_m\rbrace  $ (the set obtained from $ \mathcal{M} $ by erasing matrix $ M_j $) to have a $ q_j $-permutation structure; this construction will ensure the set to be proper. More in detail, for all $ j\!=\!1,\dots ,m $ we enforce the existence of a $ q_j $-partition  $ \Omega^{q_j}\!=\dot{\bigcup}_{i=1}^{q_j}\Omega^{q_j}_i $ of $ [n] $ on which, for all $ k\neq j $, the matrix $ M_k $ has to have a block-permutation structure.
%such thatfor all $ k\neq j $ the matrix $ M_k $ has a permutation structure on $\Omega_{q_j}  $. 
This request means that for every $ k=1,\dots ,m $ and for every $ j\neq k $ there must exist a permutation $ \sigma^k_j \!\in\! S_{q_j}$ such that for all $ i\!=\!1,\dots ,q_j $ and $ l\neq \sigma^k_j(i) $, $ M_k[\Omega_i^{q_j},\Omega_l^{q_j}] $ is a zero-matrix (see Definition \ref{def2}).
\\The main idea of the algorithm is to initialize every entry of each matrix to $ 1 $ and then, step by step, to set to $ 0 $ the entries that are not compatible with the conditions that we are requiring. %imposing 
As our final goal is to have a set of permutation matrices with the desired properties, at every step we need to make sure that each matrix dominates at least one permutation matrix, despite the increasing number of zeros among their entries. 

\begin{definition}
Given a matrix $ M$ and a $ q$-partition $ \Omega^{q}\!=\dot{\bigcup}_{i=1}^{q}\Omega^q_i $, we say that a permutation $ \sigma \in S_{q}$ is \textit{compatible} with $ M $ and $ \Omega^{q}$ if
%We say that a matrix $ M_k $ is \textit{compatible} with a $ q_j $-partition $ \dot{\bigcup}_{i=1}^{q_j}\Omega^j_i $ and a permutation $ \sigma^k_j \in S_{q_j}$ if 
for all $i=1,\dots ,q $, there exists a permutation matrix $ Q_i $ such that
\begin{equation}\label{eq:compatible}
M\bigl[\Omega^q_i, \Omega^q_{\sigma(i)}\bigr]\geq Q_i  .
\end{equation} 
\end{definition}

The algorithm itself is formally presented in Listing 1; we here describe in words how it operates. 
Each entry of each matrix is initialized to $ 1 $. The algorithm has two for-loops: the outer one on $ j=1,\dots ,m $, where a $ q_j $-partition $ \Omega^{q_j}=\dot{\bigcup}_{i=1}^{q_j}\Omega^{q_j}_i $ of $ [n] $ is uniformly randomly sampled, and the inner one on $k=1,\dots,m$ with $ k\neq j $ where we verify whether there exists a permutation $ \sigma^k_j \in S_{q_j}$ that is compatible with $ M_k $ and $ \Omega^{q_j}$. If it does exist, we choose one among all the compatible permutations and the algorithm moves to the next step $ k+1 $. %The way we choose this permutation will have a big impact on the outcome of the algorithm and it will determine what we call \textit{method 2} and \textit{method 3}, that will be described later in this paragraph. 
If such permutation does not exist, then the algorithm exits the inner for-loop and it
% permutation $ \sigma^k_j \in S_{q_j} $ is uniformly randomly sampled and it is verified if $ M_k $ is compatible with $\Omega_{q_j}  $ and $ \sigma^k_j$. If it is compatible, the algorithm moves to the next step $ k+1 $.
 %If it is not compatible, the algorithm continues to randomly select another permutation $\sigma'^k_j \in S_{q_j} $ and to check the compatibility of $ M_k $ with it and $ \Omega_{q_j} $. If after $ T2 $ steps this compatibility has not been established, the algorithm exits the inner for-loop and it 
selects another $ q_j $-partition of $ [n] $; it then repeats the inner for-loop for $k=1,\dots,m$ with $ k\neq j $ with this new partition. If after $ T1 $ steps it is choosing a different $ q_j $-partition $ \bar{\Omega}^{q_j} $, the existence for each $ k\neq j $ of a permutation 
$\bar{\sigma}^k_j \in S_{q_j} $ that is compatible with $ M_k $ and $\bar{\Omega}^{q_j}  $ is not established, %qui
we stop the algorithm and we say that \emph{it did not converge}. If the inner for-loop is completed, then for each $ k\neq j $ the algorithm modifies the matrix $ M_k $ by keeping unchanged each block $ M_k\bigl[\Omega^{q_j}_i, \Omega^{q_j}_{\sigma_j^k(i)}\bigr] $ for $ i=1,\dots ,q_j $ and by setting to zero all the other entries of $ M_k $, where $ \sigma_j^k $ is the selected compatible permutation; the matrix $ M_k $ has now a block-permutation structure over the partition $\Omega^{q_j}  $. The algorithm then moves to the next step $ j+1 $. If it manages to finish the outer for-loop, we have a set of binary matrices with the desired block-permutation structures. We then just need to select a permutation matrix $ P_k\leq M_k $ for every $ k=1,\dots ,m $ and then to randomly change a $ 0 $-entry of the matrices into a $ 1 $ without modifying the block-permutation structures of the matrix: this is always possible as the blocks of the partitions are nontrivial and a permutation matrix has just $ n  $ positive entries. We finally check whether the set is primitive.\\
Here below we present the procedures that the algorithm uses:
\begin{enumerate}
\item $ [p,P]=Extractperm(M,met) $\\
This is the key function of the algorithm, formally presented in Listing 2. It returns $p\!=\!1$ if the matrix $M$ dominates a permutation matrix, it returns $p\!=\!0$ and $ P\!=\!M $ otherwise. In the former case it also returns a permutation matrix $P$ selected among the ones dominated by $M$ according to $met$; if $met=2$ the matrix $P$ is sampled uniformly at random, if $met=3$ we make the choice of $P$ deterministic. More in detail,
%returns a permutation matrix dominated by matrix \verb!M!; there will be always at least one due to the use of subprocedure \verb!DomPerm!. For this procedure we implemented an algorithm where we can change a parameter, represented by \verb!met!, to make the sample of \verb!P! \emph{uniform} among all the permutation matrices that are dominated by \verb!M! or to make it deterministic. 
the procedure works as follows: we first count the numbers of ones in each column and in each row of the matrix $M$. We then consider the row or the column with the least number of ones; if this number is zero we stop the procedure and we set $p=0$, as in this case $M$ does not dominate a permutation matrix. Otherwise, we choose one of the $ 1 $-entries of the row or the column attaining this minimum: if $met=2$ (\textbf{method 2}) the entry is chosen uniformly at random while if $met=3$ (\textbf{method 3}) we take the first $ 1 $-entry in the lexicographic order. Suppose that the chosen entry is in position $ (i,j) $: we set to zero all the other entries in row $ i $ and column $ j $ and we iterate the procedure on the submatrix obtained from $M$ by erasing row $ i $ and column $ j $.
We can prove that this procedure is well-defined and in at most $ n $ steps it produces the desired output. %: $p=0$ if and only if $M$ does not dominate a permutation matrix and, in case $p=1$, method 2 indeed sample uniformly one of the permutations dominated by $M$, while method 3 is deterministic and the permutation obtained usually has its $ 1 $s distributed around the main diagonal. 
Method 3 will play an important role in our numerical experiments in Section \ref{sec:numexp} and in the discovery of new families of automata with quadratic reset threshold in Section \ref{sec:famaut}.

\item $[a,A]=DomPerm(M,\Omega,met)$\\
It returns $a\!=\!1$ if there exists a permutation compatible with the matrix $M$ and the partition $\Omega=\dot{\bigcup}_{i=1}^{q}\Omega^q_i$, it returns $a\!=\!0$ and $A\!=\!M$ otherwise. In the former case it chooses one of the compatible permutations $ \sigma $ according to $met$ and returns the $ n\times n $ matrix $A$ such that $A[\Omega^q_i, \Omega^q_{\sigma(i)}]= M[\Omega^q_i, \Omega^q_{\sigma(i)}]$ for all $ i=1,\dots ,q $, and all the other entries of $ A $ are equal to zero. %equal to $ M $ but the entries not in the blocks defined 
%returns \verb!a=1! if the matrix \verb!M! is compatible with the partition \verb!Omega! and the permutation \verb!sigma! and in this case it returns the matrix \verb!A! equal to \verb!M! but the entries not in the blocks defined 
%by (\ref{eq:compatible}) are set to zero; 
$A$ has then a block-permutation structure on $\Omega$. %Details of this procedure can be found in the Appendix.
More precisely, $DomPerm$ acts in two steps: it first defines a $ q\times q $ matrix $ B $ such that, for all $ i,k=1,\dots ,q $,\\
$\qquad \qquad \qquad
B[i,k]=\begin{cases}
1 & \text{if } M[\Omega_i^q,\Omega_k^q] \text{ dominates a permutation matrix }\\
0 & \text{ otherwise}
\end{cases}
;$\\this is done by calling $ExtractPerm$ with input $ M[\Omega_i^q,\Omega_k^q] $ and $met$ for all $ i,k=1,\dots ,q $. Notice that there exists a permutation compatible with $M$ and $\Omega$ if and only if $ B $ dominates a permutation matrix, so the second step of the procedure is to call again $[p,P]=ExtractPerm(B,met)$: if $p=0$ we set $a=0$ and $A=M$, while if $p=1$ we set $a=1$ and $A$ as described before with $\sigma=P$ (i.e.\ $ \sigma(i)=j $ iff $ P[i,j]=1 $).%; indeed the permutation $P$ is one of the permutations compatible with $M$ and $\Omega$.

\item $M\!set=Addone(P_1,...,P_m)$\\
It changes a $ 0$-entry of one of the matrices $ P_1,...,P_m $ into a $ 1 $ preserving all its block-permutation structures. The matrix and the entry are chosen uniformly at random and the procedure iterates the choice till it finds a compatible entry (which always exists); it then returns the final perturbed permutation set $M\!set$.
\item $pr=Primitive(M\!set)$\\
It returns $pr\!=\!1$ if the matrix set $M\!set\!=\! \lbrace M_1,\dots ,M_m\rbrace $ is primitive and $pr\!=\!0$ otherwise. It first verifies if the set is irreducible by checking the strong connectivity of the digraph $D_N$ where $N\!=\!\sum_{i=1}^k M_i$ (see Section \ref{sec:def}) via breadth-first search on every node, %using Tarjan's algorithm (see for example \cite{Duff}), %which has $ O(n) $ time complexity on average and $ O(n^2) $ time in the worst case. 
then primitivity is checked by the Protasov-Voynov algorithm (\cite{ProtVoyn}, Section 4).% which has in our case a time-complexity of $ (4m+1)n^2 $.
\end{enumerate}
 %in $ O(\sqrt{n}m) $, with $ n $ the number of vertices of the bipartition and $ m $ the number of edges; in our case 
%$ (2ms+1)n^2 $ with $ s $ the maximun number of positive entries of the columns of the matrices of $ \mathcal{M} $; in our case, the computational cost is 
All the above routines have polynomial time complexity in $ n $, apart from routine $Primitive$ that has time complexity of $ O(mn^2) $.
\begin{remark}\label{rem:alpin}
\begin{enumerate}
\item In all our numerical experiments the algorithm always converged, i.e.\ it always ended before reaching the stopping value $ T1 $, for $ T1 $ large enough. This is probably due to the fact that the matrix dimension $ n $ grows exponentially as the number of matrices $ m $ increases, which produces enough degrees of freedom. We leave the proof of this fact for future work. 

\item A recent work of Alpin and Alpina (\cite{Alpin}, Theorem 3) generalizes Theorem \ref{thmProt} for the characterization of primitive sets %, where the sets are required to be NZ and irreducible, 
to sets that are allowed to be reducible and the matrices to have zero columns (but not zero rows). Clearly, DFAs fall within this category.
%Without going into many details (for which we refer the reader to \cite{Alpin}, Theorem 3), Alpin and Alpina show that an $ n $-state automaton is \textit{not} synchronizing if and only if there exist a partition $ \dot{\bigcup}_{j=1}^s\Omega_j $ of $ [n] $ such that it has a block-permutation structure on a \textit{subset} of that partition. This characterization is clearly less restrictive: 
 %it just suffices to find a subset $ I\subset [s] $ such that for each letter $ A $ of the automaton there exists a permutation $ \sigma\in S_I $ such that for all $ i\in I $, if $ A(i,j)=1 $ then $ j\in \Omega_{\sigma(i)} $. 
Our algorithm could leverage this recent result in order to directly construct proper synchronizing DFAs.  We also leave this for future work.
\end{enumerate}
\end{remark}

\begin{lstlisting}[caption={Algorithm for generating proper primitive sets.},label=list:8-6,captionpos=t,abovecaptionskip=-\medskipamount]
Input: q_1,...,q_m,T1,met
Initialize M_1,...,M_m as all-ones matrices
for j:=1 to m do 
	t1=0
	while t1<T1 do	
		t1=t1+1
		choose a q_j-partition Omega_j
		for k=1 to m and k!=j do
		   [a,A_k]=DomPerm(M_k,Omega_j,met)
		   if a==0, exit inner for-loop  end
		end
		if a==1, exit while-loop  end	
	end
	if t1==T1
	    display 'does not converge', exit procedure	
	else
       	    set M_k=A_k for all k=1,...,m and k!=j
	end
end
for i:=1 to m do
    [p_i,P_i]=Extractperm(M_i,met)
end
Mset=Addone(P_1,...,P_m)
pr=Primitive(Mset)
return Mset, pr
\end{lstlisting}

\begin{lstlisting}[caption={Procedure for extracting a permutation matrix from a binary one.},label=list,captionpos=t,abovecaptionskip=-\medskipamount]
Input: M, met
n= size of M
P=M, p=1, I=[1,2,..,n], J=[1,2,..,n]
for i:=1 to n do
  v1= vector of the number of 1s in the rows of P indexed by I
  v2= vector of the number of 1s in the columns of P indexed by J
  v=[v1,v2]
  sort v in ascending order
  if v(1)==0
    p=0, P=M, exit procedure   
  else
    if v(1) belongs to v1
	 choose a 1-entry in row v(1) according to met
	 j= column index of the 1-entry chosen 
	 set to 0 all the other entries in row v(1) and column j
	 delete v(1) from I, delete j from J
    else
	 choose a 1-entry in column v(1) according to met
	 i= row index of the 1-entry chosen
         set to 0 all the other entries in column v(1) and row i
	 delete v(1) from J, delete i from I
    end    
  end
end
return p, P
\end{lstlisting}

\subsection{Numerical results}\label{sec:numexp}

We here compare four methods of generating random primitive sets with respect to the magnitude of the reset threshold of their associated synchronizing DFAs and we show that our randomized procedure manages to generate synchronizing DFAs with quadratic reset threshold.  \\
We call \textbf{method 1} the sets generated by Procedure \ref{proc} with $ m=2 $ (two matrices); \textbf{method 2} and \textbf{method 3}, already introduced in the previous paragraph, refer to our randomized construction where, respectively, a permutation matrix is extracted from a binary one uniformly at random or deterministically. Finally, we call \textbf{method 4} a set generated by the following procedure:

\begin{procedure}\label{proc2}
\begin{enumerate}
\item Two permutation matrices $ P_1 $ and $ P_2 $ are sampled uniformly and independently at random from $ S_n $;
\item A $ 1 $-entry of $ P_1 $ is selected uniformly at random. Suppose this entry is in row $ i $ and column $ j $; we select uniformly an index $ \bar{j}\in [n]\setminus \lbrace j\rbrace $ and we set $ P_1[i,j]=0 $ and $ P_1[i,\bar{j}]=1 $;
\item  Let $ i'\neq i $ be the other index such that $ P_1[i',\bar{j}]=1 $ (it always exists as $ P_1 $ is a permutation matrix). We select uniformly an index $ \bar{i}\in [n]\setminus \lbrace i,i' \rbrace $ and we set $ P_1[\bar{i},j]=1 $. 
\end{enumerate}
\end{procedure}
The matrix $ P_1 $ generated by Procedure \ref{proc2} does \emph{not} dominate a permutation matrix and it has the least number of positive entries that an NZ-matrix that does not dominate a permutation matrix can have. Procedure \ref{proc2} has been developed because Theorem \ref{Thm:primperm} and Theorem \ref{thm:binarymat} show that when all the matrices of the set dominate a permutation matrix with high probability we expect low exponents.

\begin{figure}
\includegraphics[scale=0.21]{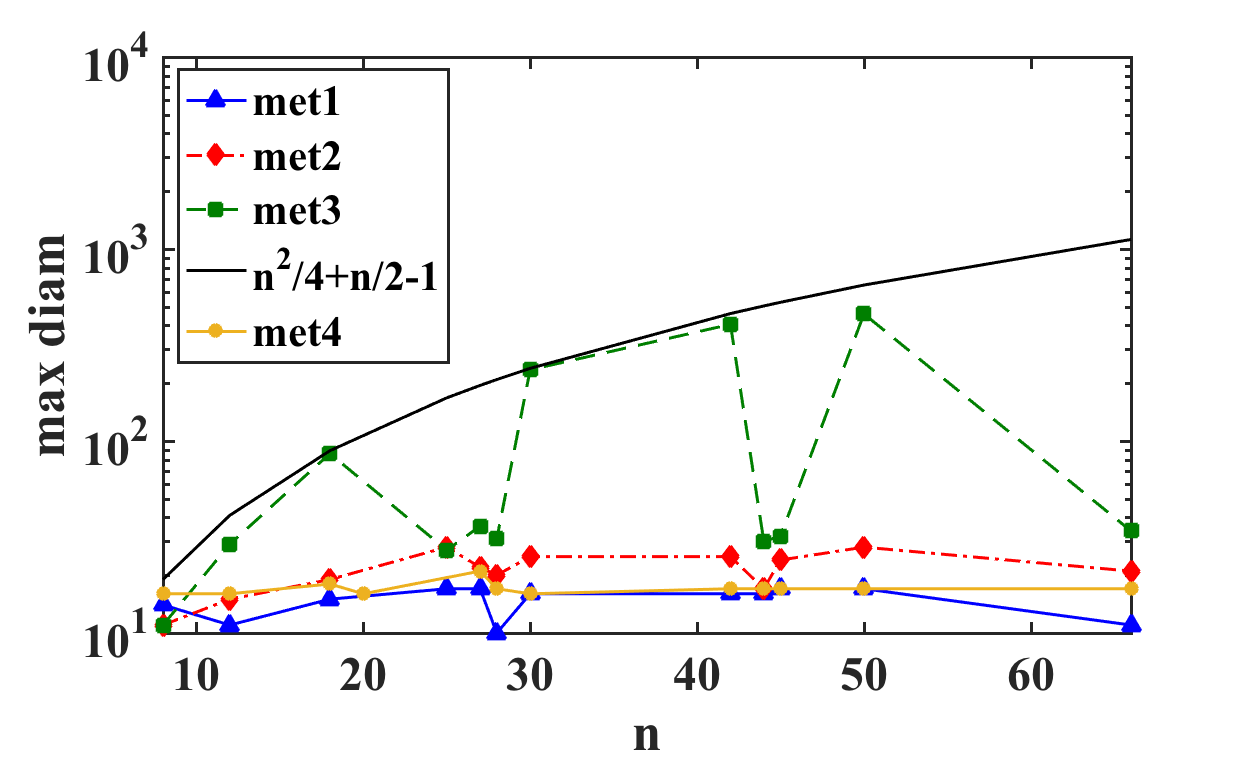}
\caption{Comparison between methods 1, 2, 3 and 4 with respect to the maximal diameter found on $ 50n^2 $ iterations when $n$ is the product of three prime numbers; the $ y $ axis is in logarithmic scale.}
\label{fig:3mat}
\end{figure}

\begin{figure}
\includegraphics[scale=0.2]{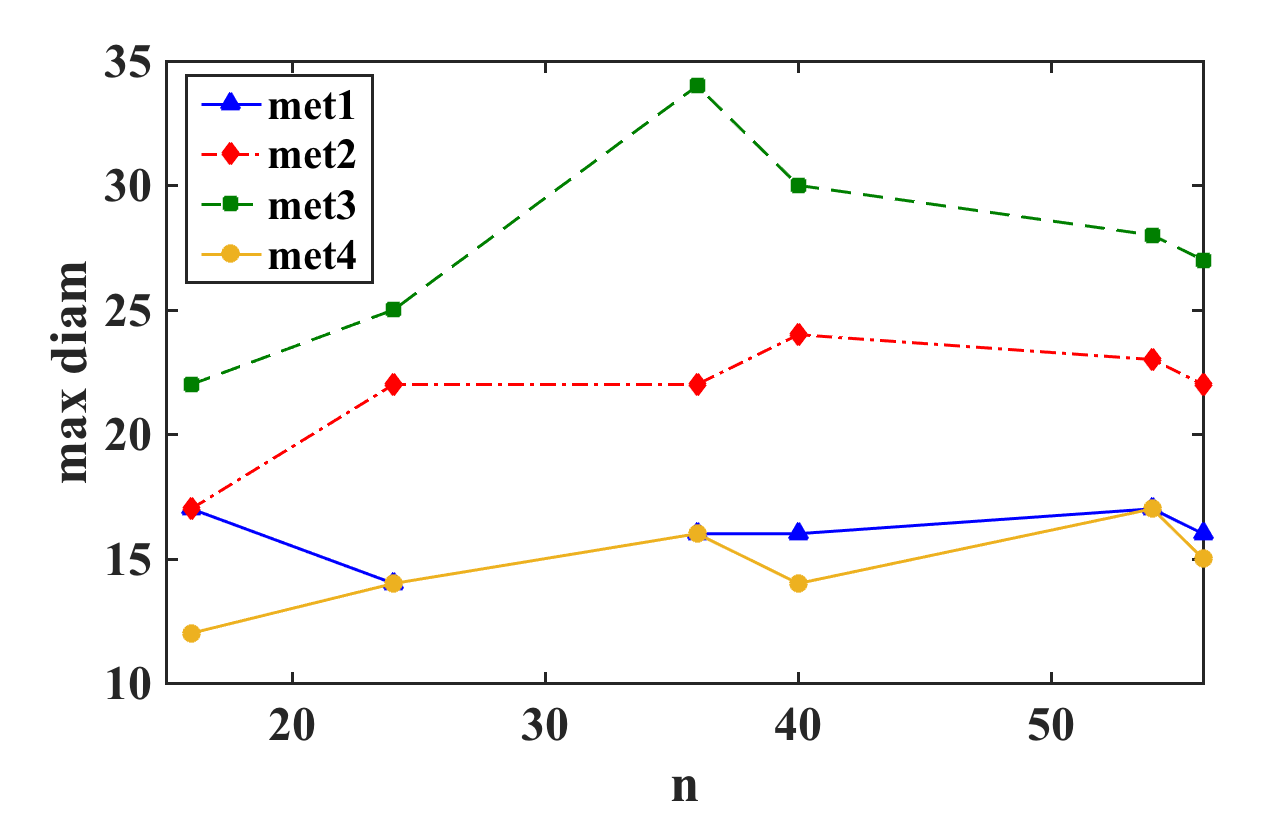}
\caption{Comparison between methods 1, 2, 3 and 4 with respect to the maximal diameter found on $ 50n^2 $ iterations when $n$ is the product of four prime numbers.}
\label{fig:4mat}
\end{figure}

\begin{figure}
\includegraphics[scale=0.2]{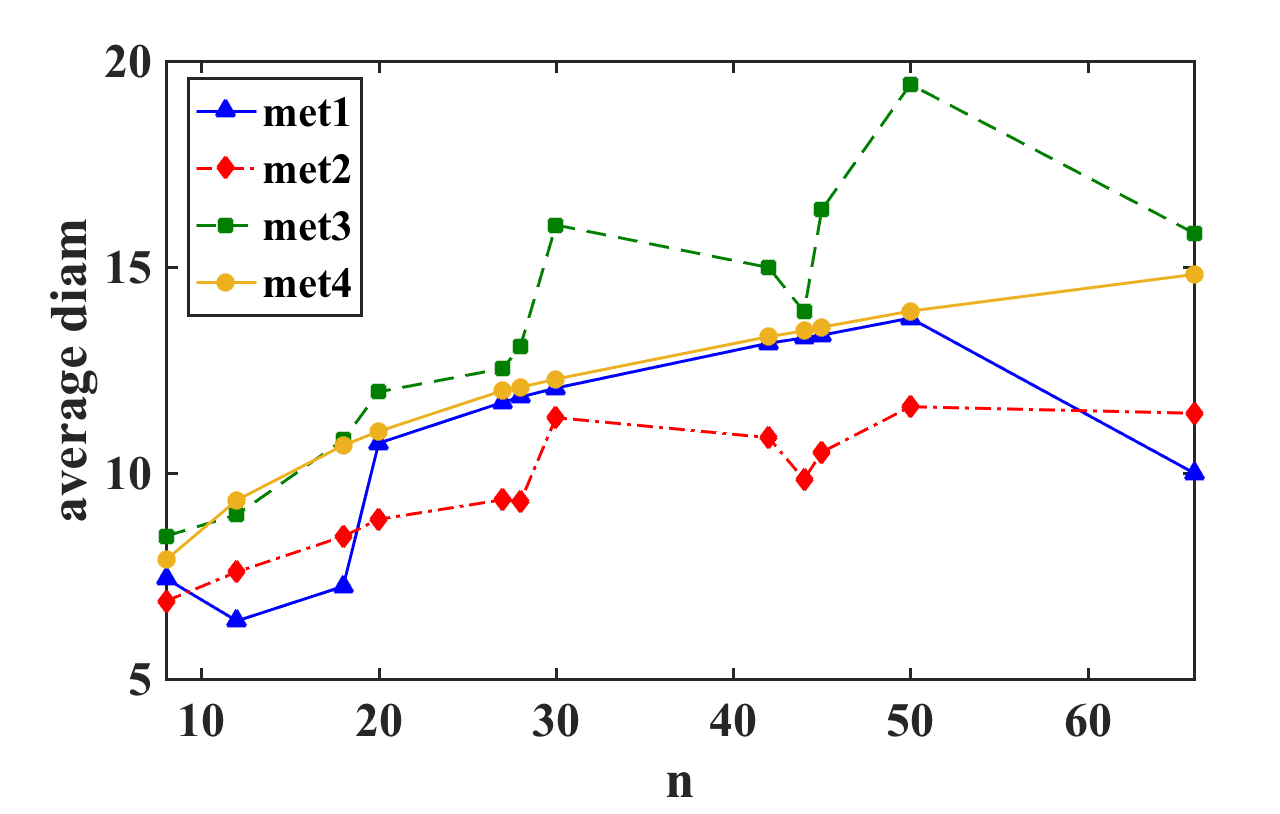}
\caption{Average diameter obtained by methods 1, 2, 3 and 4 when $n$ is the product of three prime numbers.}
\label{fig:avg}
\end{figure}

\begin{figure}
\includegraphics[scale=0.21]{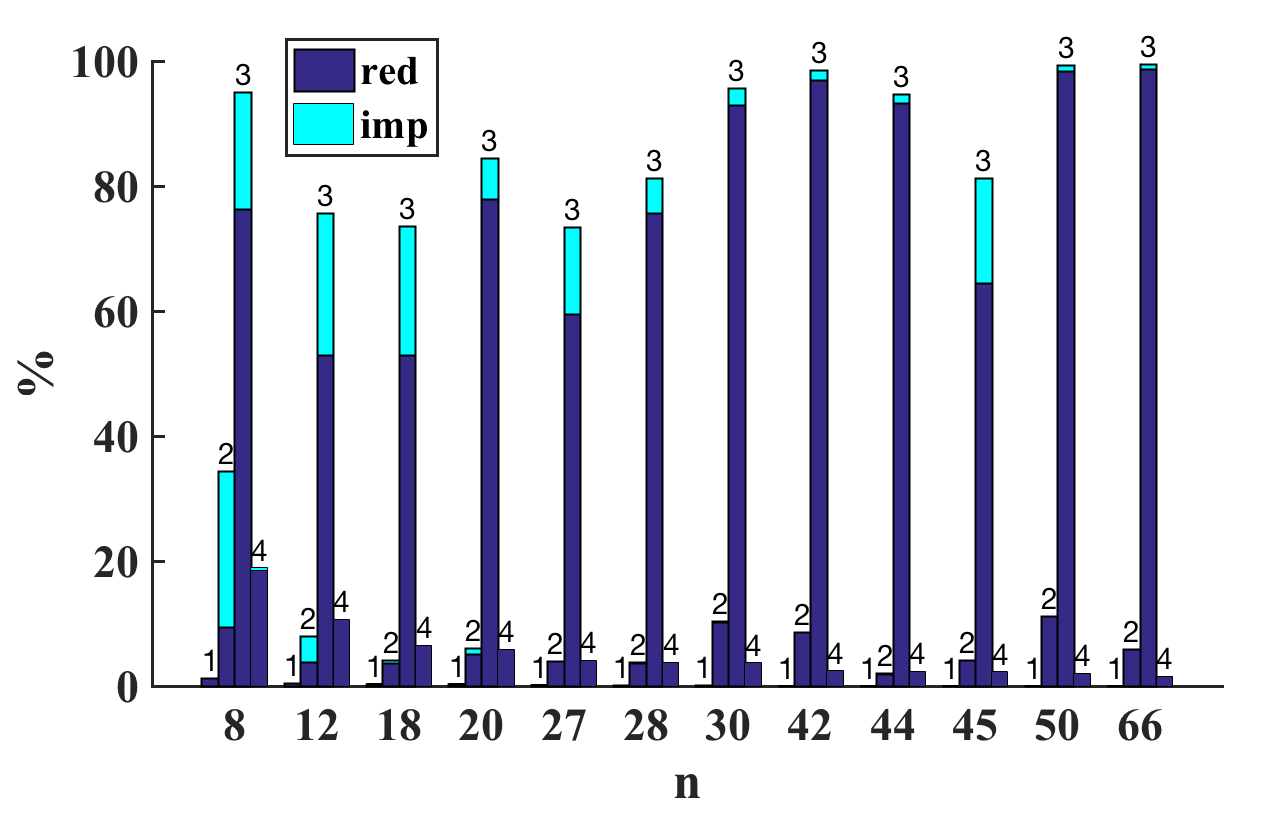}
\caption{Percentage of nonprimitive sets (divided into reducible and imprimitive sets) generated by methods 1, 2, 3 and 4 (indicated above each bar) when $n$ is the product of three prime numbers. For instance, on sets of dimension $ n=20 $, method $1$ generates $ 0.35\% $ of nonprimitive sets ($ 0.35\% $ reducible, $ 0\% $ imprimitive), method $2$ generates $ 6.15\% $ of nonprimitive sets ($ 5.18\% $ reducible, $ 0.97\% $ imprimitive), method $3$ generates $ 84.5\% $ of nonprimitive sets ($ 77.9\% $ reducible, $ 6.6\% $ imprimitive) and method $4$ generates $ 5.88\% $ of nonprimitive sets ($ 5.88\% $ reducible, $ 0\% $ imprimitive).}
\label{fig:redimp}
\end{figure}

For each method and each choice of $ n $ we run the algorithm $it(n)= 50n^2 $ times, thus producing each time $ 50n^2 $ sets.
This choice for $ it(n) $ has been made by taking into account two facts: on one hand, it is desirable to keep constant the rate $ it(n)/k_m(n) $ between the number of sampled sets $ it(n) $ and the cardinality $ k_m(n) $ of the state space. Since $ k_m(n+1)/k_m(n) $ grows approximately as $ n^m $, we have that $ k_m(n) $ explodes very fast and so we also have to deal with the limited computational speed of our computers. The choice of $it(n)= 50n^2 $ comes as a compromise between these two issues, at least when $ n\leq 70 $.
Among the $ it(n) $ generated sets, we select the primitive ones and we generate their associated DFAs (Definition \ref{def:assoc_autom}); we then check which ones are not proper synchronizing and we make them proper by using Proposition \ref{prop:minsetsautom} (it is easy to prove a similar result for method 4). We set $ T1\!=\!1000 $ for method $ 2 $ and $ 3 $.
Due to the fact that computing the reset threshold of thousands of generated instances is prohibitive (computing the reset threshold of an automaton is an NP-hard problem \cite{Epp}), we use a proxy for the reset threshold, the so called \emph{diameter of the square graph}, which is introduced here below. The square graph diameter is computable in polynomial time, namely $ O(mn^2) $ with $ m $ the number of letters of the automaton and $ n $ its number of states. 
\begin{definition}\label{def:sg}
The \emph{square graph} $ S(\mathcal{A}) $ of an $ n $-state DFA $ \mathcal{A} $ is the labeled directed graph with vertex set $ V\!=\!\lbrace (i,j): 1\leq i\leq j\leq n\rbrace $ and edge set $ E $ such that $ e\!=\!\lbrace (i,j), (i',j')\rbrace\!\in \! E $ if there exists a letter $ A\!\in\!\mathcal{A} $ such that $ A[i,i']>0 $ and $ A[j,j']>0 $, or $ A[i,j']>0 $ and $ A[j,i']>0 $. In this case, we label the edge $ e $ by $ A $ (multiple labels are allowed). A vertex of type $ (i,i) $ is called a \emph{singleton}.\\
The \emph{diameter} of $ S(\mathcal{A}) $, indicated by $ diam( S(\mathcal{A}) ) $, is the maximum of 
$ d(u,s) $ on any non-singleton vertex $ u$ and any singleton $ s $, where $ d $ denotes the length of the shortest path from $ u $ to $ s $.
\end{definition}
A well-known result (\cite{Volk}, Proposition 1) states that a DFA is synchronizing if and only if in its square graph there exists a path from any non-singleton vertex to a singleton one; the proof of this fact also implies that 
\begin{equation}\label{eq:sg}
diam\bigl(S(\mathcal{A})\bigr)\leq rt(\mathcal{A})\leq n\cdot diam\bigl(S(\mathcal{A})\bigr).
\end{equation}
The diameter of the square graph thus represents a lower bound on the reset threshold of an automaton and can be hence used as a proxy.\\%; in our numerical result we report the diameter of the square graph of the associated automata of the sets generated by methods 1, 2, 3 and 4.\\ 
%where $ diam\bigl(S(\mathcal{A})\bigr) $ denotes the \emph{diameter} of $S(\mathcal{A})  $ i.e. the maximum length of the shortest path between any two given vertices, taken over all the pairs of vertices. %In practice, it sufficies to compute the smallest length of the shortest path between a non-singleton vertex $ (i,j) $ and any singleton, and then take the maximum over all $ i\neq j $.
%The diameter can be computed in polynomial time, namely $ O(mn^2) $ with $ m $ the number of letters of the automaton. 
%\begin{remark}
%Definition \ref{def:sg} can be used for irreducible NZ-matrix sets as well. Also in this case it holds that an irreducible NZ-matrix set is primitive if and only if in its square graph there exists a path from any non-singleton vertex to a singleton one; this relies on the fact that, for an irreducible NZ-matrix set, primitivity is equivalent to admitting a product with an all-ones column (\cite{ProtVoyn}, Lemma 4). Notice that a primitive set and its associated automaton (Definition \ref{def:assoc_autom}) share the same square graph (with different labelling).
%\end{remark}
%We now report our numerical results based on the diameter of the square graph. 
%qui
Figure \ref{fig:3mat} reports on the $ y $ axis the maximal square graph diameter found among the associated automata of the sets generated by methods 1, 2, 3 and 4 for each matrix dimension $ n $ when $ n $ is the product of three prime numbers. Figure \ref{fig:4mat} reports the same but when $ n $ is the product of four prime numbers. We can see that our randomized construction manages to reach higher values of the square graph diameter than the mere random generation; in particular, method 3 reaches quadratic diameters in case of three matrices. %On the $ y $ axis is reported the maximum square graph diameter found on $ 50n^2 $ iterations by the three methods: on the left side is the case where $ n $ is a product of three primes while on the right side is the case where $ n $ is a product of four primes.  
We also report in Figure \ref{fig:avg} the behavior of the \emph{average} diameter of the proper synchronizing automata generated on $ 50n^2 $ iterations when $ n $ is the product of three prime numbers: we can see that in this case method $ 2 $ does not perform better than method $1$ and $ 4 $, while method $ 3 $ performs just slightly better. This behavior could have been expected since our primary goal was to randomly generate \emph{at least one} slowly synchronizing automata, which is indeed what happens with method $3$ that manages to reach quadratic reset thresholds most of the times.\\
A remark can be done on the percentage of the generated sets that are \emph{not} primitive; this is reported in Figure \ref{fig:redimp}, where we divide nonprimitive sets into two categories: reducible sets and \emph{imprimitive} sets, i.e.\ irreducible sets that are not primitive. We can see that the percentage of nonprimitive sets generated by method 1 and 4 goes to $ 0 $ as $ n $ increases, behavior that we partially expected (see Section \ref{sec:whp}, Theorem \ref{Thm:primperm}), while method 2 seems to always produce a non-negligible percentage of nonprimitive sets, although quite small. The behavior is reversed for method $ 3 $: most of the generated sets are not primitive. This can be interpreted as a good sign. Indeed, nonprimitive sets can be seen as sets with \emph{infinite} exponent; as we are generating a lot of them with method $3$, we intuitively should expect that, when a primitive set is generated, it has high chances to have large diameter.\\
%indeed, we know that for any $ n\times n $ primitive set $ \mathcal{M} $ it holds that $ exp(\mathcal{M})\leq (n^3+2n-3)/3 $ (\cite{BlonJung}, Corollary 18) and moreover, if the \v{C}ern\'{y} conjecture holds, then $ exp(\mathcal{M})\leq 2n^2-3n+1 $ (by Theorem \ref{thm:autom_matrix}). On the other hand, we can think about nonprimitive sets as sets whose exponent is infinite. Therefore primitive sets with quadratic exponent can be seen as the boundary between \emph{fast} primitive sets (primitive sets with small exponent) and nonprimitive sets. 
%Sampling sets according to a distribution that is concentrated around this boundary, as method 3 seems to do, should intuitively lead to a greater probability to generate sets with large exponent. \\
The slowly synchronizing automata found by our randomized construction are presented in the following section. 

\section{New families of synchronizing DFAs with quadratic reset threshold}\label{sec:famaut}
We present here four new families of slowly synchronizing automata with square graph diameter of order $ \Omega(n^2/4)$, which represents a lower bound for their reset threshold; they have been found via the randomized algorithm described in Section \ref{sec:algorithm} via method 3.
Our families are made of proper synchronizing automata with three letters: two symmetric permutation matrices and a matrix that fixes all the states but one. This characteristic makes our families differ from the \v{C}ern\'{y} automaton and the other known families of extremal automata (e.g.\ \cite{SlowAutom,BondtDon,GusevPriba,GusevSzikulaDzyga,Szykula2015}) to the extent that, if we set $ r(\mathcal{A})=\min\lbrace k\in\mathbb{N}: a^k=a, \,\,\forall \, a\in\Sigma\rbrace $ where $ \Sigma $ is the alphabet of the DFA $ \mathcal{A} $, then $ r(\mathcal{A})=3$ for any automaton $ \mathcal{A} $ that belongs to our families while $ r(\mathcal{C}_n)=n+1$ for the \v{C}ern\'{y} automaton on $ n $ states $ \mathcal{C}_n $\footnote{And similarly $r(\mathcal{A})$ is linear in $ n $ for most of the known extremal automata.}. %In particular, our automata are made of two symmetric permutations matrices and a perturbed identity matrix, which make them %Our families have been found via the randomized algorithm described in Section \ref{sec:algorithm} via method 3; they also 
Our families belong to the class of automata \textit{with simple idempotents} introduced by Rystsov in \cite{Rystsov}, who proved an upper bound of $ 2(n-1)^2 $ on their reset threshold,
and they are the associated DFAs of primitive sets made of a perturbed identity matrix and two symmetric permutations. The following proposition shows that primitive sets of this kind must have a very specific shape. With a slight abuse of notation we identify a permutation matrix $ Q $ with its underlying permutation, that is we say that $ Q(i)=j $ if and only if $ Q[i,j]=1 $; the identity matrix is denoted by $ I $. Note that a permutation matrix is symmetric if and only if its cycle decomposition is made of fixed points and cycles of length $ 2 $. 
\begin{proposition}\label{prp:auotmashape}
Let $ \mathcal{M}_{ij}=\lbrace \bar{I}_{ij}, Q_1, Q_2\rbrace $ be a matrix set of $ n\times n $ matrices where $\bar{I}_{ij}=I+\mathbb{I}_{ij}$, $ j\neq i $, is a perturbed identity and $ Q_1 $ and $ Q_2 $ are two symmetric permutations. If $ \mathcal{M} $ is irreducible then, up to a relabeling of the vertices, $ Q_1 $ and $ Q_2 $ have the following form:\\
- if $ n $ is even
\small
\begin{equation}\label{saus1}
Q_1(i)=\begin{cases}
1 &\text{if } i=1\\
i+1 &\text{if } i \text{ even, }2\leq i\leq n-2\\
i-1 &\text{if } i \text{ odd, }3\leq i\leq n-1\\
n &\text{if } i=n\\
\end{cases},\quad
Q_2(i)=\begin{cases}
i-1 &\text{if } i \text{ even}\\
i+1 &\text{if } i \text{ odd}\\
\end{cases}
\end{equation}
\normalsize
or
\small
\begin{equation}\label{saus2}
Q_1(i)=\begin{cases}
n &\text{if } i=1\\
i+1 &\text{if } i \text{ even, }2\leq i\leq n-2\\
i-1 &\text{if } i \text{ odd, }3\leq i\leq n-1\\
1 &\text{if } i=n\\
\end{cases},\quad
Q_2(i)=\begin{cases}
i-1 &\text{if } i \text{ even}\\
i+1 &\text{if } i \text{ odd}\\
\end{cases}
\end{equation}\normalsize
- if $ n $ is odd
\small
\begin{equation}\label{sausodd}
Q_1(i)=\begin{cases}
1 &\text{if } i=1\\
i+1 &\text{if } i \text{ even}\\
i-1 &\text{if } i \text{ odd, }3\leq i\leq n\\
\end{cases},\quad
Q_2(i)=\begin{cases}
i-1 &\text{if } i \text{ even}\\
i+1 &\text{if } i \text{ odd, }1\leq i\leq n\!-\!2\\
n &\text{if } i=n\\
\end{cases}.
\end{equation}
\end{proposition}

\begin{proof}
The set $ \mathcal{M} $ is irreducible if and only if the digraph $ D $ induced by matrix $ \bar{I}_{ij}+Q_1+Q_2 $ is strongly connected (see Section \ref{sec:def}). If $D$ is strongly connected, then the digraph induced by $ Q_1+Q_2 $ must be strongly connected as $ Q_1 $ and $ Q_2 $ are symmetric and the matrix $ \bar{I}_{ij} $ adds just a single edge that is not a selfloop in $ D $.
%: indeed, if not, it means that it has two connected components $ \mathcal{C}_1 $ and $ \mathcal{C}_2 $ with $ i\in\mathcal{C}_1 $ and $ j\in\mathcal{C}_2 $ connected in $ D $ by the direct edge $ (i,j) $, the only edge induced by $ \bar{I} $ that is not a loop. As this is the only edge that can connect $ \mathcal{C}_1 $ to $ \mathcal{C}_2$, there is no direct path from any vertex of $ \mathcal{C}_2$ to any vertex of $ \mathcal{C}_1$ and so $ D $ would not be strongly connected. 
Consider vertex $ 1 $: there must exist a matrix in the set $ \lbrace Q_1,Q_2\rbrace $ that links it to another vertex; let this matrix be $ Q_2 $ (wlog) and label this vertex with $ 2 $. As $ Q_2 $ is symmetric, we have $ Q_2(1)=2 $ and $ Q_2(2)=1 $. This implies that $ Q_1 $ needs to link vertex $ 2 $ to some vertex other than $ 1 $ as otherwise the digraph would not be strongly connected; we label this vertex with $ 3 $ and so we have $ Q_1(2)=3 $ and $ Q_1(3)=2 $. By iterating this reasoning, it follows that $ Q_1 $ and $ Q_2 $ must be as in (\ref{saus1}) or (\ref{saus2}) if $ n $ is even or as in (\ref{sausodd}) if $ n $ is odd. 
\end{proof}

\begin{proposition}\label{prop:noprimset}
A matrix set $ \mathcal{M}_{ij}=\lbrace \bar{I}_{ij}, Q_1, Q_2\rbrace $ of type (\ref{saus2}) is never primitive.
\end{proposition}

\begin{proof}
Due to the symmetry of digraph $ D_{Q_1+Q_2} $, up to a relabeling of the vertices we can assume without loss of generality that $ i\!=\! 1 $. If $ j $ is odd, all the three matrices have a block-permutation structure over the partition $\left\lbrace \lbrace 1,3,\dots ,n-1\rbrace,\lbrace 2,4,\dots ,n\rbrace \right\rbrace $, while if $ j $ is even they have a block-permutation structure over the partition 
$
\bigl\lbrace \lbrace 1,k\rbrace ,\lbrace 2,k-1\rbrace ,\dots ,\lbrace \frac{k}{2},\frac{k}{2}+1\rbrace, \lbrace k+1,n\rbrace , \lbrace k+2,n-1\rbrace , \dots $  $
\dots ,\lbrace \frac{n+k}{2},\frac{n+k}{2}+1\rbrace\bigr\rbrace .
$
By Theorem \ref{thmProt}, the set cannot be primitive. 
\end{proof}

We now present our new families of slowly synchronizing automata, prove closed formulas for their square graph diameter and finally state a conjecture on their reset thresholds.

%We say that a binary row-stochastic matrix $ A $ is \textit{idempotent} if $ A^2 =A$. Rystsov proved that for the class of automata made up of just idempotent and permutation matrices, also called automata \textit{with simple idempotent}, there exists a quadratic upperbound on the rest threshold:
%\begin{theorem}[Theorem 3 in \cite{Rystsov}]
%Every synchornizing $ n $-state automata with simple idempotent has reset threshold less or equal than $ 2(n-1)^2 $.
%\end{theorem}
%The bound above is twice the bound stated by the \v{C}ern\'{y} conjecture, but indeed, a priori, the reset threshold of this family could have a much smaller growth rate than quadratic. 

\begin{definition}\label{defn:Aij}
Let $ \mathcal{M}_{ij}=\lbrace \bar{I}_{ij}, Q_1, Q_2\rbrace $ where $\bar{I}_{ij}=I+\mathbb{I}_{ij}$ for $ j\neq i $ and $ Q_1 $ and $ Q_2 $ are as in eq.(\ref{saus1}) if $ n $ is even and as in eq.(\ref{sausodd}) if $ n $ is odd.
We define $ \mathcal{A}_{ij}=\lbrace \underline{I}_{ij}, Q_1,Q_2\rbrace $ to be the associated DFA (see Definition \ref{def:assoc_autom}) of $ \mathcal{M}_{ij} $, where $ \underline{I}_{ij}=I+\mathbb{I}_{ij}-\mathbb{I}_{ii} $.
\end{definition}

Figure \ref{fig:autom} represents the automaton $ \mathcal{A}_{1,6} $ with $ n=8 $. We set $ \mathcal{E}_n=\mathcal{A}_{1,n-2} $ for $ n=4k $ and $ k\geq 2 $, $ \mathcal{E}'_n=\mathcal{A}_{1,n-4} $ for $ n\!=\!4k+2 $ and $ k\geq 2 $, $ \mathcal{O}_n=\mathcal{A}_{\frac{n-1}{2},\frac{n+1}{2}} $ for $ n=4k+1 $ and $ k\geq 1 $, $ \mathcal{O}'_n=\mathcal{A}_{\frac{n-1}{2},\frac{n+1}{2}} $ for $ n=4k+3 $ and $ k\geq 1 $. The following theorem holds:

\begin{figure}
\begin{tikzpicture}[shorten >=1pt,node distance=2.5cm,on grid,auto,scale=0.45,transform shape,inner sep=0.8pt,bend angle=60,every state/.style={text=black},every node/.style={text=black}]
  %\draw[help lines] (0,0) grid (3,2);
  \node[state]  (q_0)                      {1};
  \node[state]          (q_1) [ right=of q_0] {2};
  \node[state]          (q_2) [right =of q_1] {3};
  \node[state](q_3) [ right=of q_2] {4};
  \node[state](q_4) [ right=of q_3] {5};
  \node[state](q_5) [ right=of q_4] {6};
  \node[state](q_6) [ right=of q_5] {7};
  \node[state](q_7) [ right=of q_6] {8};
  \path[->, line width=0.3mm] (q_0) edge [bend left]				node        {} (q_5);        
   \path[->, dotted,line width=0.2mm ](q_0) edge [bend left]				node        {} (q_1)
  (q_2) 	edge [bend left]	           		node  {} (q_3)
  (q_3) edge [bend left]				node        {} (q_2)
  (q_1) 	edge [bend left]	           		node  {} (q_0)
  (q_4) edge [bend left]				node        {} (q_5)
  (q_6) 	edge [bend left]	           		node  {} (q_7)
  (q_7) edge [bend left]				node        {} (q_6)
  (q_5) 	edge [bend left]	           		node  {} (q_4) ;
\path[->] (q_0) edge [loop left]			node  {} ()	
          (q_1) edge [bend left]				node        {} (q_2)
		(q_3) 	edge [bend left]	           		node  {} (q_4)
		(q_4) edge [bend left]				node        {} (q_3) 
		(q_2) 	edge [bend left]	           		node  {} (q_1)		
(q_5) edge [bend left]				node        {} (q_6)
		(q_7) 	edge [loop right]			node  {} () 	 
		(q_6) 	edge [bend left]	           		node  {} (q_5)
		;	
\end{tikzpicture}
\caption{The automata $ \mathcal{A}_{1,6} $ with $ n=8 $; $ rt(\mathcal{A}_{1,6}) $=$31  $. Dashed arrows refer to matrix $ Q_2 $, normal arrows to matrix $ Q_1 $ and bold arrows to matrix $ \underline{I}_{1,6} $, where its selfloops have been omitted.}\label{fig:autom}
\end{figure}
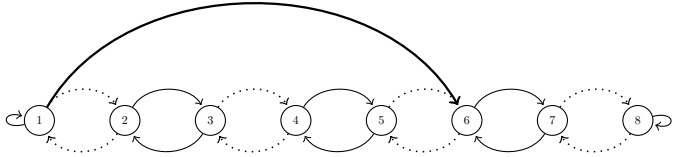

\begin{theorem}\label{thm:autfam}
The automaton $ \mathcal{E}_n$ has square graph diameter (SGD) of $ (n^2+2n-4)/4 $, $ \mathcal{E}'_n$ has SGD of $ (n^2+2n-12)/4 $, $ \mathcal{O}_n$ has SGD of $ (n^2+3n-8)/4 $ and $ \mathcal{O}'_n$ has SGD of $ (n^2+3n-6)/4$. Therefore all the families $ \mathcal{E}_n$, $ \mathcal{E}'_n$, $ \mathcal{O}_n$ and $ \mathcal{O}'_n$ have reset threshold of $ \Omega(n^2/4) $.
\end{theorem}

\begin{figure}
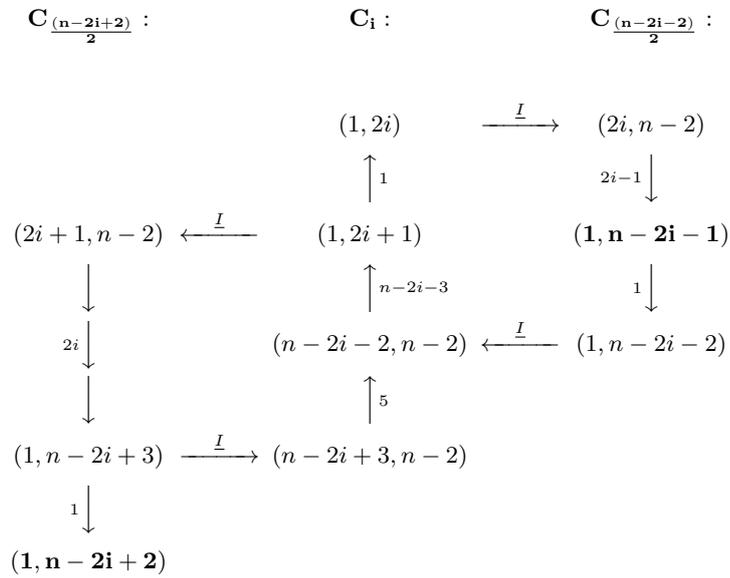

\begin{align*}
\small
\begin{CD}
\mathbf{C_{\frac{(n-2i+2)}{2}}} :      @.    \mathbf{C_i}: @. \mathbf{C_{\frac{(n-2i-2)}{2}}}: \\
      @.     @. \\
      @.   (1,2i)       @>\underline{I}>> (2i,n-2)\\
      @.     @AA{1}A       @V{2i-1}VV \\
      (2i+1,n-2)  @<\underline{I}<<   (1,2i+1) @.     \mathbf{(1,n-2i-1)} \\
    @VVV       @AA{n-2i-3}A       @V{1}VV \\
   @V{2i}VV        (n-2i-2,n-2)       @<\underline{I}<< (1,n-2i-2) \\
   @VVV			@AA{5}A				@.\\
  (1,n-2i+3)  @>\underline{I}>> (n-2i+3,n-2) @.\\
   @V{1}VV  @. @.\\
  \mathbf{ (1,n-2i+2)}@. @.\\
\end{CD}
\end{align*}
\caption{Diagram on how the components $ \lbrace C_i\rbrace $ in the proof of Theorem \ref{thm:autfam}  are linked together. Vertices in the same column belong to the same component (indicated above the column).}\label{fig:diagram}
\end{figure}

\begin{figure}
\begin{tikzpicture}[shorten >=1pt,node distance=1.3cm,on grid,auto,scale=0.5,transform shape,inner sep=0.4pt,bend angle=70,every state/.style={text=black},every node/.style={text=black}]

\node at (0,6) {\huge{$ \mathbf{ C_0}$}};
\node at (5,6) {\huge{$\mathbf{ C_1 }$}};
\node at (10,6) {\huge{$ \mathbf{C_{\frac{n-4}{2}}} $}};
\node at (14,6) {\huge{$ \mathbf{\dots} $}};
\node at (19,6) {\huge{$\mathbf{ C_{\frac{n}{2}-1} }$}};

 % \draw[help lines] (0,0) grid (25,6);
  \node[state] [red]  (q_4)                      {$\frac{n}{2}$, $\frac{n}{2}$+1};
  \node[state]          (q_3) [ above=of q_4] {3,n-2};
  \node[state]          (q_2) [above=of q_3, shape =rectangle] {$ \vdots $};
   \node[state]          (q_1) [above=of q_2] {1,n};

  \node[state](q_5) [ right=of q_1 , xshift=3.6cm] {};
  \node[state](q_6) [ right=of q_2, xshift=3.6cm] {n-2,n};
  \node[state](q_7) [ right=of q_3, xshift=3.6cm] {};
   \node[state](q_8) [ right=of q_4, xshift=3.6cm] {n-4,n-2};
  \node[state](q_9) [ below=of q_8,shape =rectangle ] {$ \vdots $};
 % \node[state](q_10) [ below=of q_9] {};
  \node[state](q_11) [ below=of q_9] {1,3};
  \node[state](q_12) [ below=of q_11] {1,2};
  
  \node[state](q_13) [ right=of q_5 , xshift=3.6cm] {};
  \node[state](q_14) [ right=of q_6, xshift=3.6cm] {1,n-4};
  \node[state](q_15) [ right=of q_7, xshift=3.6cm] {1,n-3};
   \node[state](q_16) [ right=of q_8, xshift=3.6cm] {2,n-2};
   \node[state](q_17) [ right=of q_9 , xshift=3.6cm,shape =rectangle] {$ \vdots $};
  %\node[state](q_18) [ below=of q_17, xshift=3.6cm] {};
  %\node[state](q_19) [ right=of q_11, xshift=3.6cm] {};
   \node[state](q_20) [ right=of q_11, xshift=3.6cm] {n-2,n-1};
   
 \node[state](q_30) [ right=of q_16, xshift=3cm, shape =rectangle] {$\cdots $};
 %\node[state](q_31) [ right=of q_30, xshift=1.8cm, shape =rectangle] {$\cdots\cdots\cdots  $};

\node[state](q_21) [ right=of q_13 , xshift=8cm] {};
  \node[state](q_22) [ right=of q_14, xshift=8cm,shape =rectangle] {$\vdots $};
  \node[state](q_23) [ right=of q_15, xshift=8cm] {5,n-2};
   \node[state](q_24) [ right=of q_16, xshift=8cm,shape =rectangle] {$\vdots $};
   \node[state](q_25) [ right=of q_17 , xshift=8cm] {1,n-1 };
  \node[state](q_26) [ below=of q_25] {1,n-2 };
  \node[state](q_27) [ below=of q_26,shape =rectangle] {$\vdots $};
   \node[state](q_28) [ below=of q_27] {$ \frac{n}{2}$-1,$\frac{n}{2} $};

   \node[state](q_29) [ right=of q_21 , xshift=2cm] {n-2,n-2};
   
 \path[->, line width=0.3mm] (q_1) edge[red] node {} (q_6)
 (q_11) edge			node        {} (q_3)	
  (q_12)  edge[red]				node        {} (q_16)
  (q_14) edge 			node        {} (q_8)	
(q_15)  edge	[red]		node        {} (q_30)	
  (q_26) edge	[red]		node        {} (q_29)
  (q_30) edge				node        {} (q_20)
	(q_30) edge	[red]			node        {} (q_26)
	(q_25) edge			node        {} (q_30)
  ;
  \path[->, dotted,line width=0.2mm ]
  (q_1) edge [bend left]				node        {} (q_2)
  (q_2) edge [bend left,red]			node        {} (q_1)	
  (q_3) edge [bend left]				node        {} (q_4)
  (q_4) edge [bend left,red]				node        {} (q_3)	
  (q_5) edge [loop above ]				node        {} ()
	(q_6) edge [bend left,red]				node        {} (q_7)	
	(q_7) edge [bend left]				node        {} (q_6)
	(q_8) edge [bend left,red]			node        {} (q_9)	
	(q_9)edge [bend left]				node        {} (q_8)	
	(q_9) edge [bend left,red]				node        {} (q_11)
	(q_11) edge [bend left]				node        {} (q_9)	
	(q_12) edge [loop below ]				node        {} ()
	(q_13) edge [bend left]				node        {} (q_14)
	(q_14) edge [bend left]				node        {} (q_13)
	(q_15) edge [bend left]				node        {} (q_16)
	(q_16) edge [bend left,red]			node        {} (q_15)
	(q_17) edge [bend left]				node        {} (q_20)
	%(q_18) edge [bend left]				node        {} (q_17)
	%(q_19) edge [bend left]				node        {} (q_20)
	(q_20)	edge [bend left]				node        {} (q_17)
	(q_21) edge [loop above ]				node        {} ()
	(q_22) edge [bend left]				node        {} (q_23)
	(q_23) edge [bend left]				node        {} (q_22)
	(q_24) edge [bend left]				node        {} (q_25)
	(q_25) edge [bend left]				node        {} (q_24)	
	(q_26) edge [bend left]				node        {} (q_27)
	(q_27) edge [bend left]				node        {} (q_26)
	(q_28) edge [loop below ]				node        {} ()	
  ;
\path[->] (q_1) edge [loop above ]				node        {} ()						
(q_2) edge [bend left]				node        {} (q_3)	
(q_3) edge [bend left,red]				node        {} (q_2)			
(q_4) edge [loop below ]				node        {} ()		
(q_5) edge [bend left]				node        {} (q_6)
(q_6) edge [bend left]				node        {} (q_5)	
(q_7) edge [bend left,red]				node        {} (q_8)	
(q_8) edge [bend left]				node        {} (q_7)	
%(q_9) edge [bend left]				node        {} (q_10)	
%(q_10) edge [bend left]				node        {} (q_9)	
(q_11) edge [bend left,red]				node        {} (q_12)
(q_12) edge [bend left]				node        {} (q_11)
(q_13) edge [loop above ]				node        {} ()				
(q_14) edge [bend left]				node        {} (q_15)
(q_15) edge [bend left]				node        {} (q_14)	
(q_16) edge [bend left]				node        {} (q_17)
(q_17) edge [bend left]				node        {} (q_16)	
%(q_18) edge [bend left]				node        {} (q_19)
%(q_19) edge [bend left]				node        {} (q_18)	
(q_20) edge [loop below ]				node        {} ()	
(q_21) edge [bend left]				node        {} (q_22)
(q_22) edge [bend left]				node        {} (q_21)	
(q_23) edge [bend left]				node        {} (q_24)
(q_24) edge [bend left]				node        {} (q_23)	
(q_25) edge [bend left]				node        {} (q_26)	
(q_26) edge [bend left]				node        {} (q_25)		
(q_27) edge [bend left]				node        {} (q_28)	
(q_28) edge [bend left]				node        {} (q_27)	
					;	
\end{tikzpicture}
\caption{Square graph of the family $ \mathcal{E}_{n} $, where all the singletons but $ (n-2,n-2) $ have been omitted. There are $ n/2 $ chains, the first one ($ \mathbf{C_0} $) has $ n/2 $ vertices, the others have $ n $ vertices; the missing chains and vertices are represented by squared boxes with dots. Normal lines refer to matrix $ Q_1 $, dotted lines to matrix $ Q_2 $ and bold lines to matrix $ \underline{I}$, where its selfloops have been omitted. The red path is the diameter. }\label{figSG}
\end{figure}
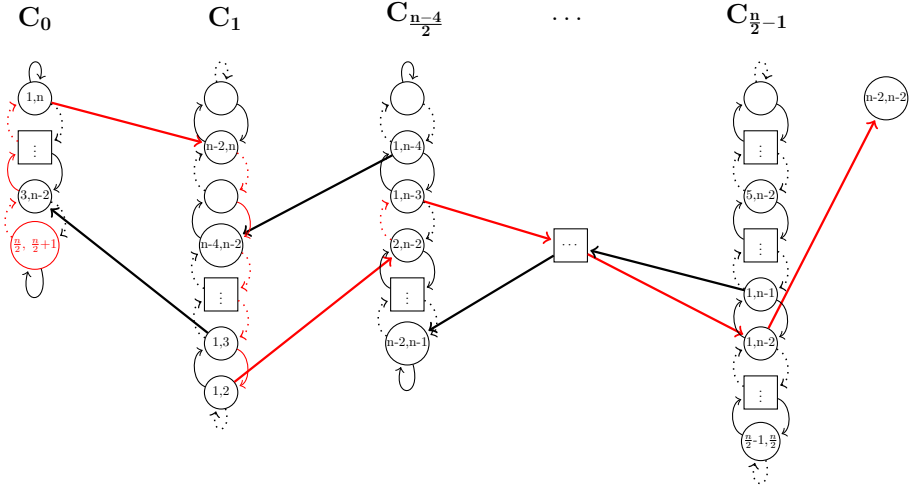

\begin{figure}[h!]
\begin{tikzpicture}[shorten >=1pt,node distance=1.3cm,on grid,auto,scale=0.45,transform shape,inner sep=0.4pt,bend angle=70,every state/.style={text=black},every node/.style={text=blue}]
  %\draw[help lines] (0,0) grid (3,2);
  \node[state] [red] (q_4)                      {4,5};
  \node[state]          (q_3) [ above=of q_4] {3,6};
  \node[state]          (q_2) [above=of q_3] {2,7};
   \node[state]          (q_1) [above=of q_2] {1,8};

  \node[state](q_5) [ right=of q_1 , xshift=3.6cm] {7,8};
  \node[state](q_6) [ right=of q_2, xshift=3.6cm] {6,8};
  \node[state](q_7) [ right=of q_3, xshift=3.6cm] {5,7};
   \node[state](q_8) [ right=of q_4, xshift=3.6cm] {4,6};
  \node[state](q_9) [ below=of q_8 ] {3,5};
  \node[state](q_10) [ below=of q_9] {2,4};
  \node[state](q_11) [ below=of q_10] {1,3};
  \node[state](q_12) [ below=of q_11] {1,2};
  
  \node[state](q_13) [ right=of q_5 , xshift=3.6cm] {2,3};
  \node[state](q_14) [ right=of q_6, xshift=3.6cm] {1,4};
  \node[state](q_15) [ right=of q_7, xshift=3.6cm] {1,5};
   \node[state](q_16) [ right=of q_8, xshift=3.6cm] {2,6};
   \node[state](q_17) [ right=of q_9 , xshift=3.6cm] {3,7};
  \node[state](q_18) [ right=of q_10, xshift=3.6cm] {4,8};
  \node[state](q_19) [ right=of q_11, xshift=3.6cm] {5,8};
   \node[state](q_20) [ right=of q_12, xshift=3.6cm] {6,7};
   
% \node[state](q_30) [ right=of q_16, xshift=3cm, shape =rectangle] {$\cdots $};
 %\node[state](q_31) [ right=of q_30, xshift=1.8cm, shape =rectangle] {$\cdots\cdots\cdots  $};

\node[state](q_21) [ right=of q_13 , xshift=3.6cm] {3,4};
  \node[state](q_22) [ right=of q_14, xshift=3.6cm] {2,5};
  \node[state](q_23) [ right=of q_15, xshift=3.6cm] {1,6};
   \node[state](q_24) [ right=of q_16, xshift=3.6cm] {1,7};
   \node[state](q_25) [ right=of q_17 , xshift=3.6cm] {2,8};
  \node[state](q_26) [ right=of q_18, xshift=3.6cm] {3,8};
  \node[state](q_27) [ right=of q_19, xshift=3.6cm] {4,7};
   \node[state](q_28) [ right=of q_20, xshift=3.6cm] {5,6};

   \node[state](q_29) [ right=of q_21 , xshift=2cm] {6,6};
   
 \path[->, line width=0.3mm] (q_1) edge[red] node {} (q_6)
 (q_11) edge			node        {} (q_3)	
  (q_12)  edge[red]				node        {} (q_16)
  (q_14) edge 			node        {} (q_8)	
(q_15)  edge	[red]		node        {} (q_23)	
  (q_23) edge[red]			node        {} (q_29)
  %(q_30) edge				node        {} (q_18)
	%(q_30) edge				node        {} (q_23)
	(q_24) edge				node        {} (q_20)
  ;
  \path[->, dotted,line width=0.2mm ]
  (q_1) edge [bend left]				node        {} (q_2)
  (q_2) edge [bend left,red]				node        {} (q_1)	
  (q_3) edge [bend left]				node        {} (q_4)
  (q_4) edge [bend left,red]				node        {} (q_3)	
  (q_5) edge [loop above ]				node        {} ()
	(q_6) edge [bend left,red]				node        {} (q_7)	
	(q_7) edge [bend left]				node        {} (q_6)
	(q_8) edge [bend left,red]				node        {} (q_9)	
	(q_9)edge [bend left]				node        {} (q_8)	
	(q_10) edge [bend left,red]				node        {} (q_11)
	(q_11) edge [bend left]				node        {} (q_10)	
	(q_12) edge [loop below ]				node        {} ()
	(q_13) edge [bend left]				node        {} (q_14)
	(q_14) edge [bend left]				node        {} (q_13)
	(q_15) edge [bend left]				node        {} (q_16)
	(q_16) edge [bend left,red]				node        {} (q_15)
	(q_17) edge [bend left]				node        {} (q_18)
	(q_18) edge [bend left]				node        {} (q_17)
	(q_19) edge [bend left]				node        {} (q_20)
	(q_20)	edge [bend left]				node        {} (q_19)
	(q_21) edge [loop above ]				node        {} ()
	(q_22) edge [bend left]				node        {} (q_23)
	(q_23) edge [bend left]				node        {} (q_22)
	(q_24) edge [bend left]				node        {} (q_25)
	(q_25) edge [bend left]				node        {} (q_24)	
	(q_26) edge [bend left]				node        {} (q_27)
	(q_27) edge [bend left]				node        {} (q_26)
	(q_28) edge [loop below ]				node        {} ()	
  ;
\path[->] (q_1) edge [loop above ]				node        {} ()						
(q_2) edge [bend left]				node        {} (q_3)	
(q_3) edge [bend left,red]				node        {} (q_2)			
(q_4) edge [loop below ]				node        {} ()		
(q_5) edge [bend left]				node        {} (q_6)
(q_6) edge [bend left]				node        {} (q_5)	
(q_7) edge [bend left,red]				node        {} (q_8)	
(q_8) edge [bend left]				node        {} (q_7)	
(q_9) edge [bend left,red]				node        {} (q_10)	
(q_10) edge [bend left]				node        {} (q_9)	
(q_11) edge [bend left,red]				node        {} (q_12)
(q_12) edge [bend left]				node        {} (q_11)
(q_13) edge [loop above ]				node        {} ()				
(q_14) edge [bend left]				node        {} (q_15)
(q_15) edge [bend left]				node        {} (q_14)	
(q_16) edge [bend left]				node        {} (q_17)
(q_17) edge [bend left]				node        {} (q_16)	
(q_18) edge [bend left]				node        {} (q_19)
(q_19) edge [bend left]				node        {} (q_18)	
(q_20) edge [loop below ]				node        {} ()	
(q_21) edge [bend left]				node        {} (q_22)
(q_22) edge [bend left]				node        {} (q_21)	
(q_23) edge [bend left]				node        {} (q_24)
(q_24) edge [bend left]				node        {} (q_23)	
(q_25) edge [bend left]				node        {} (q_26)	
(q_26) edge [bend left]				node        {} (q_25)		
(q_27) edge [bend left]				node        {} (q_28)	
(q_28) edge [bend left]				node        {} (q_27)	
					;	
\end{tikzpicture}
\caption{Square graph of automaton $ \mathcal{E}_{8} $, $diam\bigl(S(\mathcal{E}_{8})\bigr)=  19$. Normal lines refer to matrix $ Q_1 $, dotted lines to matrix $ Q_2 $ and bold lines to matrix $ \underline{I}_{1,6}$, where its selfloops have been omitted. The red path is the diameter.}\label{figSG8}
\end{figure}
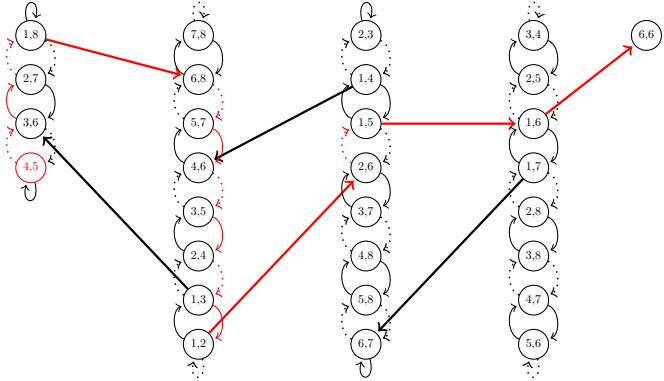

\begin{proof}
%We invite the reader to refer to the automaton in Figure \ref{figSG} through out the proof. 
We prove the theorem just for the family $ \mathcal{E}_n$; the other square graph diameters can be obtained by a similar reasoning. We set $ \underline{I}=\underline{I}_{1,n-2} $ to ease the notation. In the following we describe the shape of $ S(\mathcal{A}_{1,n-2} )$ with $ n=4k $ in order to compute its diameter, i.e. the maximal distance between a non-singleton vertex and the singleton $ (n-2,n-2) $, as it is the only singleton that has an in-going edge starting from a non-singleton vertex; we invite the reader to refer to Figure \ref{figSG} during the proof. %just the singleton $ (n-2,n-2) $ is pictured.
%In its description we omit the singleton vertices as the only link in $ S(\mathcal{A}_{1,n-2} )$ between a non-singleton vertex and a singleton one is the edge connecting $ (1,n-2) $ to $ (n-2,n-2) $. %It is clear that we can focus just on the shortest paths of $ G_{\mathcal{A}} $ connecting a non singleton vertex to the singleton $ (n-2,n-2) $. 
 %If we do not consider the merging letter $ \underline{I} $, t
The digraph $S(\mathcal{A}_{1,n-2}\setminus \lbrace \bar{I}\rbrace ) $, without considering the singletons, is disconnected and has $ n/2 $ strongly connected components: $ C_0 $ of size $ n/2 $ and $ C_1,\dots ,C_{n/2-1} $ of size $ n$. The component $ C_0 $ is made of the vertices $ \lbrace (1+s, n-s): s=0,\dots ,n/2 -1\rbrace $ while component $ C_i $ is made of the vertices $ \lbrace (i,i+1),(i-1,i+2),\dots ,(1,2i),(1,2i+1),(2,2i+2),\dots ,(n-i,n-i+1)\rbrace $ for $ 1\leq i \leq n/2-1 $: these components look like ``chains'' due to the symmetry of $ Q_1 $ and $ Q_2 $ (see Figure \ref{figSG}). In particular, the vertices $ (1,n) $ and $ (3,n-2) $ belong to $ C_ 0$, the vertices $ (1,2i)$ and $(1,2i+1) $ belong to $ C_i $ for $ 1\leq i\leq n/2\!-\!1 $, the vertices $ (n-4,n-2)$ and $(n-2,n) $ belong to $ C_1 $, the vertices $ (1,n-2)$ and $(4,n-2) $ belong to $ C_{n/2-1} $ and the vertices $ (n-2i-2,n-2)$ and $(n-2i+3,n-2) $ belong to $ C_i $ for $ 2\leq i\leq n/2\!-\!2 $. The matrix $ \underline{I}$ connects the components $\lbrace C_i\rbrace_i$ by linking vertex $ (1,a) $ to vertex $ (a,n-2) $ for every $a=2,\dots ,n$ in such a way that the $ \lbrace C_i\rbrace_i $ can be ordered from the farthest to the closest to the singleton $ (n-2,n-2) $ (see Figure \ref{figSG}). Indeed, the diagram in Figure \ref{fig:diagram} shows how the components $ \lbrace C_i\rbrace_i $ are linked together for $  2\leq i \leq n/2-1 $: an arrow between two vertices means that there exists a word mapping the first vertex to the second one, a number next to the arrow represents the length of such word if the two vertices belong to the same component while arrows connecting vertices from different components are labeled by $ \underline{I} $; bold vertices represent the ones that are linked by $ \underline{I} $ to other chains. How $ C_0 $ is connected to $ C_1 $ is directly shown in Figure \ref{figSG}. It follows that the digraph $ S(\mathcal{A}_{1,n-2}) $ is formed by ``layers'' represented by the components $\lbrace C_i\rbrace_i $ where
\begin{equation}\label{eq:sequence}
C_0,\,C_1,\,C_{\frac{n-4}{2}},\,C_3,\,C_{\frac{n-8}{2}},\,C_5,\,C_{\frac{n-12}{2}},\dots
\end{equation}
is the sequence of components from the farthest to the closest to the singleton $ (n-2,n-2) $.
%$ C_0 $ is the farthest from the singleton $ (n-2, n-2) $ and $ C_{n/2-1}$ is the closest, as shown in Figure \ref{figSG}. 
In order to compute the diameter we need to measure the length of the shortest path from vertex $ (n/2,n/2+1) $ to vertex $ (n-2,n-2) $, which is colored in red in Figure \ref{figSG}. This means that for $ 0\leq i\leq n/2-1 $ we have to compute the distance $ d_i $ in $ C_i$ between vertices $ (2i,n-2) $ and $ (1,n-2i-1) $ if $ i $ is odd or between vertices $ (2i+1,n-2) $ and $ (1,n-2i+2) $ if $ i $ is even.
In view of (\ref{eq:sequence}), %the diagram in Figure \ref{fig:diagram}, as the sequence of components from the farthest to the closest to the singleton $ (n-2,n-2) $ is $ C_0,C_1,C_{\frac{n-4}{2}},C_3,C_{\frac{n-8}{2}}, C_5,C_{\frac{n-12}{2}},\dots  $,
we have the following sequence for the $ d_i $s:
\[
d_0\!=\!\frac{n}{2}-1,\,d_1\!=\! n-2,\,d_{\frac{n-4}{2}}\!=\!1,\,d_3\!=\! n-3, \,d_{\frac{n-8}{2}}\!=\!5,\,d_5\!=\!n-7,\,d_{\frac{n-12}{2}}\!=\!9, \dots
\]
%Observe that $ d_{2i}+d_{2i+1}= n-2 $ for $ 1\leq i\leq n/4 -1 $. 
Since the number of edges labeled by $ \underline{I} $ that appear in the path is $ n/2 $, the diameter is equal to
\[
diam(S(\mathcal{A}_{1,n-2} ))=\frac{n}{2}+ \sum_{k=0}^{\frac{n}{2}-1} d_k= \frac{n^2}{4}+\frac{n}{2}-1 .
\]\end{proof}
Figure \ref{figSG8} represents the square graph of the automaton $ \mathcal{E}_{8} $, where its diameter is colored in red. All the singletons but the one that belongs to the diameter have been omitted. 

\begin{conjecture}
The automaton $ \mathcal{E}_n$ has reset threshold of $ (n^2-2)/2 $, $ \,\mathcal{E}'_n$ has reset threshold of $ (n^2-10)/2 $ and $ \mathcal{O}_n$ and $ \mathcal{O}'_n$ have reset threshold of $ (n^2-1)/2 $. Furthermore, they represent the automata with the largest possible reset threshold among the family $ \lbrace\mathcal{A}_{ij}\rbrace_{i\neq j} $ for respectively $ n=4k $, $ n=4k+2 $, $ n=4k+1 $ and $ n=4k+3 $.
\end{conjecture}
Notice that, although the randomized construction for proper primitive sets presented in Section \ref{sec:algorithm} is defined just when the matrix size $ n $ is the product of at least three prime numbers, we here presented an extremal $ n $-state automaton of quadratic reset threshold for \emph{any} value of $ n $.
%\textcolor{blue}{In \cite{Szykula2016}, Szyku{\l}a and Vorel presented a family $ \mathcal{A}_n $ of slowly eulerian synchronizing $ n $-state automata, for $ n=4k+1 $, with reset threshold of $ (n^2-3)/2 $. Our automaton $ \mathcal{O}_n $ is a sub-automaton of $ \mathcal{A}_n $, i.e.\ $ \mathcal{A}_n=\mathcal{O}_n \cup \lbrace\underline{I}_{\frac{n+1}{2},\frac{n-1}{2}}\rbrace $\footnote{We remind that $ \underline{I}_{ij}=I+\mathbb{I}_{ij}-\mathbb{I}_{ii} $.}. We have hence proved that their family $ \mathcal{A}_n $ is not proper and that their construction can be generalized to any $ n $.}\\
Theorem \ref{thm:autfam} can also be seen as an improvement in the direction initiated by Gonze et.\ al.\ in \cite{Gonze}, where they show that the square graph diameter of any $ n $-state automaton made of $ m\geq 2 $ permutation matrices is lower bounded by $ n^2/4+o(n^2) $ when $ n $ is odd. We have proved that this lower bound holds for any $ n $ and any $ n $-state \emph{synchronizing} automaton containing $ m\geq 2 $ permutation matrices.

%\begin{proposition}
%The automata family has reset threshold smaller than $ n^2-3n+3 $.
%\end{proposition}
%\begin{proof}
%Set $ \underline{I}_{1,n-2}=\underline{I} $. The word $ w=  \underline{I}\bigl( (Q_2Q_1)^{\frac{n}{2}-1}\underline{I}\bigr)^{n-2}$ is a synchronizing word of length $ n^2-3n+3 $.
%\end{proof}
\section{Conclusion}
In this paper we have exploited the connection between primitive sets of NZ-matrices and synchronizing DFAs to propose a randomized construction for generating slowly synchronizing automata. We have first shown that random perturbed permutation sets have small exponent most of the times, thus producing fast synchronizing DFAs. The same behavior applies to random binary sets (alias random NDFAs) where each entry of each matrix is independently set to $ 1 $ with probability $ p $; we have also shown that $ p(n)=(\log n+c)/n $ is a threshold for the property of these random sets to be primitive and to be $ 3 $-directable. In particular, an uniformly sampled NDFA of at least two letters has both a $ 2 $-directing word and a $ 3 $-directing word of length $ O(n\log n) $ with high probability.
Secondly, we have proposed a more involved randomized construction for primitive sets based on a recent characterization of NZ-primitive sets (Theorem \ref{thmProt}) and we have shown via Theorem \ref{thm:autom_matrix} that it is able to generate some synchronizing DFAs with quadratic reset threshold. Finally, we have presented four new families of DFAs with simple idempotents with reset threshold of order $ \Omega(n^2/4) $; to the best of our knowledge, this is one of the few cases where an extremal family of automata does not resemble %\footnote{To the extent that, if we set $ r(\mathcal{A})=\min\lbrace k\in\mathbb{N}: a^k=a, \,\,\forall \, a\in\Sigma\rbrace $ where $ \Sigma $ is the alphabet of the DFA $ \mathcal{A} $, then $ r(\mathcal{A})=3$ for any automaton $ \mathcal{A} $ that belongs to our family while $ r(\mathcal{C}_n)=n+1$ for the \v{C}ern\'{y} automaton on $ n $ states $ \mathcal{C}_n $.  } 
the \v{C}ern\'{y}'s one.
The \textit{primitive set approach} to synchronizing DFAs seems promising and we believe that some parameters of our construction, as the way a permutation matrix is extracted from a binary one or the way the partitions of $ [n] $ are selected, could be further improved in order to generate new families of slowly synchronizing automata; for example, we could think about selecting the ones in the procedure $Extractperm$ according to a given distribution. 
As mentioned at the end of Section \ref{sec:algorithm}, one can also apply the construction directly to automata by leveraging the recent result of Alpin and Alpina (\cite{Alpin}, Theorem 3). Finally, it would be of interest to determine how the exponent of $ \mathcal{B}_m(n, p)$ behaves when $ p $ is chosen differently for each matrix of the set, e.g.\ when $ \mathcal{B}_m(n, p)=\lbrace B_1(n,p_1),\dots ,B_m(n,p_m)\rbrace $ for $ p=\bigl(p_1(n),\dots ,p_m(n)\bigr)\in [0,1]^m $, $ n\in\mathbb{N} $.
%. Our strategy relies on a recent characterization of NZ primitive sets (Theorem \ref{thmProt}), together with a construction (Definition \ref{def:assoc_autom} and Theorem \ref{thm:autom_matrix}) allowing to build (slowly proper) synchronizing automata from (slowly proper) primitive sets. We have obtained four new families of automata with simple idempotents with reset threshold of order $ \Omega(n^2/4) $. 
%
% Provide your references in a bib-file and give the name here. See the
% sample bib-file jalc-ex.bib as an example. If a name should not be
% abbreviated after the first letter, as Gh. for Gheorghe, then write it
% like this: {\relax Gh}eorghe.
%
\biblio{biblOnprobprim}

% ...and we are done !!
\end{document}